\documentclass[manyauthors,nocleardouble,COMPASS]{cernphprep}

\usepackage{graphicx}
\usepackage{subfig}
\usepackage[percent]{overpic}
\usepackage[numbers, square, comma, sort&compress]{natbib}

\usepackage{times}
\usepackage{epsfig}
\usepackage{amssymb}
\usepackage{colordvi}
\usepackage{graphicx}
\usepackage{wrapfig,rotating}
\usepackage{amsmath}
\usepackage{hyperref} 
\usepackage{caption}

\usepackage{multicol}
\usepackage{booktabs}
\usepackage{multirow}

\def\lapproxeq{\lower .7ex\hbox{$\;\stackrel{\textstyle
<}{\sim}\;$}}
\def\gapproxeq{\lower .7ex\hbox{$\;\stackrel{\textstyle
>}{\sim}\;$}}

\def\text{~}

\def\hc2{(\hbar c)^2}

\def\r2{\langle r^2 \rangle}
\def\Q2{$Q^2\/$}

\def\r0{$\rho^{0}\/$}			% by O.G., 24/05/05
\def\xB{$x_{Bj}$}            
\def\pt2{$p_{T}^{2}\/$}			% by O.G., 30/05/05
\def\Gev2{GeV$^{2}$}			% by O.G., 30/05/05
\def\A1r{$A_{1}^{\rho}\/$}		% by O.G., 31/05/05
\def\AUT{$A_{UT}^{\rm{sin}(\phi-\phi_S)}$}

\def\NH3{NH$_3$}
\def\a2         {{\mbox{$a_{2}                                        \
$}}}

\def\a2pigamma  {{\mbox{$a_2^-\rightarrow \pi^-\gamma                \
$}}}

\RequirePackage{lineno}

\DeclareSymbolFont{letters}     {OML}{cmm}{m}{it}
\DeclareSymbolFont{symbols}     {OMS}{cmsy}{m}{n}
\DeclareSymbolFont{largesymbols}{OMX}{cmex}{m}{n}

\begin {document}

%\linenumbers

\begin{titlepage}
\PHnumber{2012--208}
\PHdate{17 July 2012}

\title{Exclusive $\rho^0$ muoproduction on transversely polarised protons and deuterons}

\Collaboration{The COMPASS Collaboration}
\ShortAuthor{The COMPASS Collaboration}

\begin{abstract}
The transverse target spin azimuthal asymmetry \AUT in hard exclusive
production of $\rho ^0$ me\-sons was measured at COMPASS by scattering
160\,GeV/$c$ muons off transversely polarised protons and deuterons.  The 
measured asymmetry is sensitive to the nucleon helicity-flip generalised
parton distributions $\it E^q$, which are related to the orbital angular
momentum of quarks in the nucleon. The \Q2, \xB~ and \pt2 dependence of
\AUT is presented in a wide kinematic range: 
$1\,({\rm GeV}/{\it c})^2 < Q^2 < 10\,
({\rm GeV}/{\it c})^2$, $0.003 <\: x_{Bj} ~< 0.3$ and 
$0.05\,({\rm GeV}/{\it c})^2 < \:p_T^2~<
0.5\, ({\rm GeV}/{\it c})^2$ for protons or $0.10\,({\rm GeV}/{\it c})^2 < \:p_T^2~ < 0.5\,
({\rm GeV}/{\it c})^2$ for deuterons. Results for deuterons are obtained
for the first time. The measured asymmetry is small in the whole
kinematic range for both protons and deuterons, which is consistent with
the theoretical interpretation that contributions from GPDs $\it
E^u$ and $\it E^d$ approximately cancel.
\end{abstract}

\vfill
\Submitted{(submitted to Nucl. Phys. B)}
\end{titlepage}

{\pagestyle{empty}
\input{Authors2012-rho0-transverse.tx}
\clearpage
}

\setcounter{page}{4}

\section{Introduction}
 \label{Sec_intro}

Hard exclusive electro- and muoproduction of mesons on nucleons has
played an important role in studies of strong interactions and recently
gained renewed interest as it allows access to generalised parton
distributions (GPDs)~\cite{Mueller, Ji1, Ji2, Radyu1, Radyu2}.  The GPDs
provide a novel and comprehensive description of the partonic structure
of the nucleon and contain a wealth of new information. In particular,
they embody both nucleon electromagnetic form factors and parton
distribution functions. Furthermore, GPDs provide a description of the
nucleon as an extended object, referred to as 3-dimensional nucleon
tomography~\cite{Burkardt1,Burkardt2,Burkardt3}, which correlates
longitudinal momenta and transverse spatial degrees of freedom of
partons. The evaluation of GPDs may for the first time provide insight
into angular momenta of quarks, another fundamental property of the
nucleon~\cite{Ji1,Ji2}.  The mapping of nucleon GPDs, which very
recently became one of the key objectives of hadron physics, requires a
comprehensive program of measuring various hard exclusive processes in a
broad kinematic range, in particular deeply virtual Compton scattering
(DVCS). Hard exclusive meson production provides
independent and complementary information.

In perturbative QCD (pQCD), there exists a general proof of
factorisation~\cite{fact} for exclusive meson production by longitudinal
virtual photons. In this case the amplitude for hard exclusive meson
leptoproduction can be factorised into a hard-scattering part and soft
parts, the latter depending on the structure of the nucleon described by
GPDs and on the structure of the meson described by its 
distribution amplitude (DA).  No similar
proof of factorisation exists for transverse virtual photons.  However,
pQCD-inspired models taking into account parton transverse momenta have been proposed
\cite{mrt,gk1,gk2}, which describe reasonably well the behaviour of the
cross sections for both longitudinal and transverse photons,
$\sigma_{L}\/$ and $\sigma _T\/$, respectively.

At leading twist, meson production is described by four types of GPDs:
$H^f$, $E^f$, $\widetilde{H}^f$, $\widetilde{E}^f$, where $f$ denotes
a quark of a given flavour or a gluon. The GPDs are functions of
$t$, $x$ and $\xi$, where $t$ is the squared four-momentum transfer to
the nucleon, $x$ the average and $\xi$ half the difference of the
longitudinal momenta carried by the struck parton in the initial and
final states. In addition, there is a scale dependence of GPDs which is
not explicitly shown here.  Depending on the quark content and the
quantum numbers of the meson, there exists sensitivity to various types
of GPDs and different quark flavours. In particular, production of
vector mesons is sensitive only to GPDs $H^f$ and $E^f$.

The GPDs attracted much attention after it was shown that the total
angular momentum of a given parton species $f$ is related to the second
moment of the sum of GPDs $H^f$ and $E^f$ via the Ji relation
\cite{Ji1}:
\begin{equation}
\label{eq:gpd_4}
J^f = \frac{1}{2}\:\lim_{t \rightarrow 0}\int^{1}_{-1} {\rm d}x\: x \left[H^{f}\left(x,\xi,t \right) + E^{f}\left(x,\xi,t \right) \right] ,
\end{equation}
which holds for any value of $\xi$. The
spin-independent cross sections for DVCS and for vector meson production by
longitudinal photons on a proton target are mostly sensitive to the
nucleon-helicity-conserving GPDs $H^f$, with GPDs $E^f$ being suppressed
in the COMPASS kinematic domain. However, the GPDs $E^{f}$ are of
special interest, as they are related to the orbital angular momentum of
quarks. They describe transitions with nucleon helicity flip, in which
orbital angular momentum must be involved due to total angular momentum
conservation. It was pointed out that the spin dependent
cross sections for DVCS on transversely polarised protons~\cite{DGPR,BMK} and 
for exclusive vector meson production by longitudinal photons on
transversely polarised nucleons~\cite{Goeke} are sensitive to 
the `elusive' nucleon
helicity-flip GPDs $E^f$. Access to GPDs $E^f$ is also possible by
measurements of the cross section for DVCS on an unpolarised neutron
target~\cite{BMK}.

Measurements of the lepton helicity dependent DVCS cross section on neutrons were
performed by the JLAB Hall A collaboration~\cite{neutron} and the
transverse target spin asymmetries for DVCS from transversely polarised
protons were measured by the HERMES experiment~\cite{dvcshermes}. Model-dependent
estimates of the total angular momenta of quarks, $J^u$ and
$J^d$, derived from the results of these measurements, indicate a large
value for $J^u$ and a value close to zero for $J^d$, in agreement with
results from lattice QCD~\cite{Lattice}.

In exclusive vector meson production on transversely polarised targets
the observable sensitive to the GPDs $E^f$ is the azimuthal asymmetry
\AUT (see Sec.~\ref{Sec_theor} for the definition). Here, the indices
$U$ and $T$ refer to the beam spin independent and transverse target
spin dependent cross section, and ${\sin(\phi-\phi_S)}$ indicates the
type of azimuthal modulation of the cross section. The GPDs $E^f$ appear
at leading twist only in this azimuthal asymmetry for vector meson
production by longitudinal photons.

The only previous measurement of \AUT for exclusive $\rho ^0$
electroproduction on transversely polarised protons was performed by the
HERMES experiment~\cite{hermes1,hermes2}. Its separate extraction for
longitudinally and transversely polarised $\rho ^0$ mesons gave values
consistent with zero. A model-dependent attempt was made~\cite{hermes1}
to extract the value of the total angular momentum $J^u$ of $u$ quarks
in the proton, although limited by large experimental uncertainties.

In this paper, we present results on \AUT for exclusive $\rho ^0$
meson muoproduction on transversely polarised protons and deuterons.
The experiment was carried out at CERN by the COMPASS collaboration
using the 160\,GeV/$c$ muon beam and a polarised target
filled either with lithium deuteride ($^{6}$LiD) or ammonia (NH$_3$) to
provide polarised deuterons or protons, respectively. 
% As will be
% detailed in the following, the measurements on polarised protons lead to
% a precision comparable to that of HERMES and cover an extended range of
% the Bjorken scaling variable $x_{Bj}$.  The measurements on polarised
% deuterons were done for the first time by COMPASS.

\section{Theoretical framework}
 \label{Sec_theor}
 
\begin{figure}[htb]
 \begin{center}
 \epsfig{figure=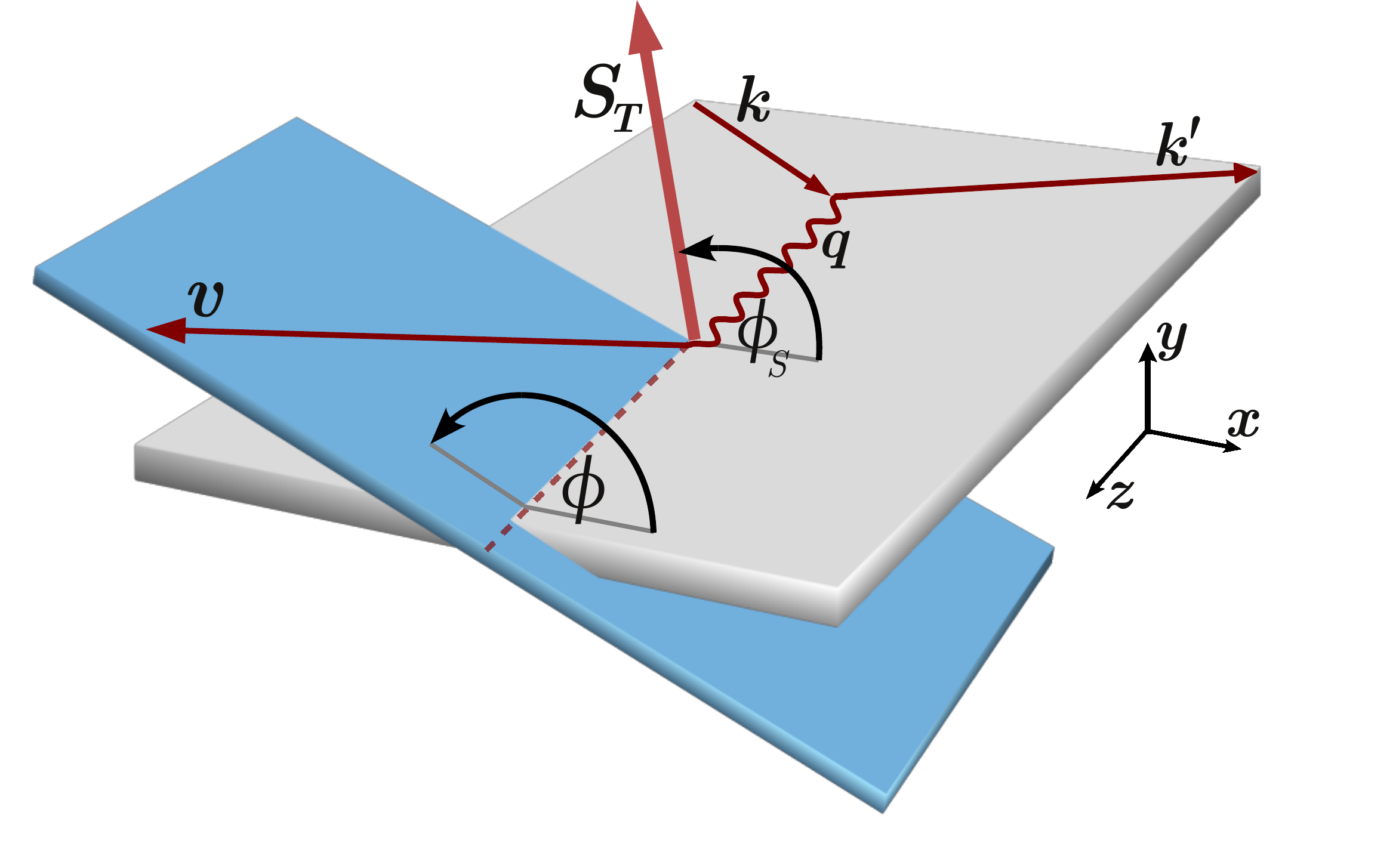,width=0.6\textwidth,angle=0,clip}

 \caption{Kinematics of exclusive meson production in the target rest frame.
Here $\pmb{k}$, $\pmb{k'}$, $\pmb{q}$ and $\pmb{v}$ represent three-momentum vectors
of the incident and the scattered muons, the virtual 
photon and the meson. $S_T$ is the component of the target spin vector $S$ 
(not shown) perpendicular to the virtual photon direction.} 
 \label{angles}
 \end{center}
\end{figure}

The cross section of hard exclusive $\rho ^0$ leptoproduction, $\mu \,N
\rightarrow \mu\, \rho^0\, N$, on a transversely polarised nucleon
depends on the photon virtuality $Q^2$, the Bjorken variable $x_{Bj}$,
$t$, $\phi$ and $\phi_{S}$~\cite{DiehlSap}. Here $\phi$ is the azimuthal
angle between the lepton scattering plane and the plane containing the
virtual photon and the produced meson (hadron plane), while $\phi_S$ is
the azimuthal angle of the target spin vector around the virtual photon
direction relative to the lepton scattering plane (see
Fig.~\ref{angles}). 
A summary of the kinematic variables used in this paper is given in 
Table~\ref{tab:intro_1}.
\begin{table}[!h]
\begin{center}
\caption{Kinematic variables.}
\label{tab:intro_1}
\vspace{0.35cm}
{\small
\begin{tabular}{l l}
\toprule
$k$                                                &  four-momentum of incident muon \\
$k'$                                               &  four-momentum of scattered muon \\
$p$                                                &  four-momentum of target nucleon \\
$v$                                                &  four-momentum of $\rho^0$ meson \\
$q=k-k'$                                           &  four-momentum of virtual photon \\
$Q^2=-q^2$                                         &  invariant negative mass squared of virtual photon \\
$W=\sqrt{(p+q)^2}$                                 &  invariant mass of the $\gamma^{*}-N$ system \\
$M_p$                                              &  proton mass \\
$\nu=(p \cdot q)/M_p$                              &  energy of virtual photon in the laboratory system \\
$x_{\mathit{Bj}}=Q^2/(2M_p \nu)$                   &  Bjorken scaling variable \\
$y=(p \cdot q)/(p \cdot k)$                        &  fraction of lepton energy lost in the laboratory system \\
$M_{\pi\pi}$                                       &  invariant mass of $\pi^+\pi^-$ system \\
$t=(q-v)^2$                                        &  square of the four-momentum transfer to the target nucleon \\
%$t'=|t|-t_{0}$                                     &  where $t_{0}$ is the minimal kinematically allowed $|t|$ for \\
%                                                   &  given $W$, $Q^2$, $M_{\pi\pi}$ and $M_X^2$ \\
$p_T^2$                                            &  transverse momentum squared of vector meson with \\
                                                   &  respect to the virtual photon direction \\
$E_{\rho^0}$                                       &  energy of $\rho ^0$ in the laboratory system \\
$M_X^2=(p+q-v)^2$                                  &  missing mass squared of the undetected system \\
$E_{\textrm{miss}}=((p+q-v)^2 - p^2)/(2M_p)$       &  missing energy of the undetected system \\
$\phantom{E_{\textrm{miss}}}=({M_X^2 - M_p^2})/({2M_p})$
                                                   &  \\
$\phantom{E_{\textrm{miss}}}=\nu - E_{\rho ^0} + t/(2 M_p)$
                                                   &  \\
%$t=(q-v)^2$                                        &  square of the four-momentum transfer to the target nucleon \\
%$t'=|t|-t_{0}$                                     &  where $t_{0}$ is the minimal kinematically allowed $|t|$ for \\
%                                                   &  given $W$, $Q^2$, $M_{\pi\pi}$ and $M_X^2$ \\
%$p_T^2$                                            &  transverse momentum squared of vector meson with \\
%                                                   &  respect to the virtual photon direction \\
%$E_{\rho^0}$                                       &  energy of $\rho ^0$ in the laboratory system \\
\bottomrule
%\multicolumn{1}{c}{\rule{0.3\textwidth}{0px}} &
%\multicolumn{1}{c}{\rule{0.6\textwidth}{0px}}
\end{tabular}
}
\end{center}
\end{table}

In the COMPASS kinematic
region the cross section can be expressed as:
%\begin{eqnarray}
\begin{align}
\label{eq:access_1}
\nonumber & \left[ \frac{\alpha_{em}}{8\pi^{3}} \frac{y^{2}}{1-\epsilon} \frac{1-x_{Bj}}{x_{Bj}} \frac{1}{Q^{2}} \right]^{-1}
 \frac{{\rm d}\sigma}{{\rm d}x_{Bj} {\rm d}Q^{2} {\rm d}t {\rm d}\phi {\rm d}\phi_{s}} \simeq  \\
& \frac{1}{2}\left( \sigma_{++}^{++} + \sigma_{++}^{--} \right) + \epsilon \sigma_{00}^{++}
- S_{T} \sin \left (\phi-\phi_{s} \right) \rm{Im} \left( \sigma_{++}^{+-} + \epsilon \sigma_{00}^{+-} \right)  + \: ... \: \: , 
%\end{eqnarray}
\end{align}
where only terms relevant for the present analysis are shown explicitly. 
%Not shown are higher-twist terms for the transversely polarised target,
%which have different azimuthal modulations, nor terms depending on the
%combination of longitudinal beam and transverse target polarisations. 
The general formula for the cross section for meson leptoproduction,
which contains the dependence on the projectile and target spins and the
complete azimuthal dependence, can be found in Ref.~\cite{DiehlSap}.
The component of the transverse target spin perpendicular to the virtual
photon direction, $S_T$, is in the COMPASS kinematic region very well
approximated by the corresponding component perpendicular to the
direction of the incoming muon.  The virtual photon polarisation
parameter $\epsilon$ is given by
\begin{equation}
\label{eq:epsilon}
\epsilon = \frac{1-y-\frac{1}{4}y^2\gamma ^2}{1-y+\frac{1}{2}y^2+\frac{1}{4}y^2\gamma ^2}, 
\end{equation}
where $y$ is the virtual-photon fractional energy, $\gamma = 2 x_{Bj} M_p/Q$
and $M_p$ the proton mass. 

The symbols $\sigma_{mn}^{ij}$ represent spin-dependent photoabsorption cross sections or interference terms, 
which are proportional to bilinear combinations of helicity 
amplitudes ${\cal{A}}^i_m$ for 
the subprocess $\gamma^{*}N \rightarrow \rho ^0N$, 
\begin{equation}
\label{eq:access_2}
\sigma_{mn}^{ij} \propto \sum_{\rm{spins}} \left( {\cal{A}}_{m}^{i} \right)^{*}{\cal{A}}_{n}^{j} ,
\end{equation}
where the dependence on kinematic variables is omitted for brevity.
The virtual-photon helicity is denoted by $m(n) = 0, \pm1$, the target
nucleon helicity by $i(j) = \pm \frac{1}{2}$, and the notation is 
restricted to $0, +, -$ for legibility.

The transverse target spin dependent cross section is accessed
experimentally by measuring the azimuthal asymmetry \AUT, which is
proportional to the ${\rm{sin}(\phi-\phi_S)}$ moment of the cross
section for a transversely polarised target. It
can be expressed as
\begin{eqnarray}
\label{eq:access5}
  A_{UT}^{\rm{sin}(\phi-\phi_S) } & = &  - \: \frac
  {
    \rm{Im}(\sigma_{++}^{+-} + \epsilon \, \sigma_{00}^{+-})
  }
  {\frac{1}{2} (\sigma_{++}^{++} + \sigma_{++}^{--}) + \epsilon \, \sigma_{00}^{++} 
  }
= - \: \frac{\rm{Im}(\sigma_{++}^{+-} + \epsilon \, \sigma_{00}^{+-})}{\sigma _T + \epsilon\, \sigma _L} \: .
\end{eqnarray}
The denominator contains the spin-averaged cross section with
contributions from both transverse and longitudinal virtual photons. The
leading twist interference term $\sigma_{00}^{+-}$ corresponds to
$\rho^0$ production by longitudinal photons, while the higher twist term
$\sigma_{++}^{+-}$ corresponds to production by transverse photons. The
former term is proportional to a weighted sum of convolutions of the
GPDs $E^{f}$ with the DA of the produced
meson and a hard scattering kernel~\cite{Goeke}. The weights depend on
the contributions of quarks of various flavours and gluons to the
production of a given vector meson.
 
The direct method to separate terms arising from production by
longitudinal and transverse photons is the Rosenbluth
separation. However, it is not feasible with the present data as only measurements
at one beam energy are available. In
Ref.~\cite{Diehl} another method was proposed which can be used for
vector meson production in the approximation of $s$-channel helicity
conservation and exploiting the decay angular distributions of the
meson.  It was applied by HERMES for the analysis of
$\rho ^0$ production on transversely polarised protons. In our analysis
we do not attempt such a separation.
Nevertheless, the present `unseparated' results can be compared to the
predictions of existing models which take into account also higher twist
effects.

\section{Experimental set-up}
 \label{Sec_exper}

The experiment~\cite{compass} was performed using the high intensity
positive muon beam from the CERN SPS M2 beam line.  The instantaneous
$\mu ^+$ beam intensity during extraction is about $4 \cdot 10^7$/s. The
average beam momentum is 160\,GeV/$c$ with a spread of 5\,GeV/$c$.  The
momentum of each incident muon is measured upstream of the experimental
area with a relative precision better than 1$\,\%$. The $\mu ^{+}$ beam
is longitudinally polarised by the weak decays of the parent
hadrons. Note that the beam polarisation does not affect the measurement
of \AUT.

The beam traverses a polarised solid-state target which contains 120 cm
total length of polarisable material, which is either 
NH$_3$ for polarised protons or $^6$LiD for
polarised deuterons. Both protons and deuterons can be
polarised either longitudinally or transversely with respect to the beam
direction.  A mixture of liquid $^3$He and $^4$He, used to refrigerate
the target, and a small admixture of other nuclei are also present in
the target. It consists of either three (NH$_3$) or two ($^6$LiD) 
separate cells with polarisable material, placed one after another along
the beam.  The spin directions in neighbouring cells are opposite. Both
target configurations allow for a simultaneous measurement of azimuthal
asymmetries for the two target spin directions to compensate flux-dependent
systematic uncertainties.  In order to reduce systematic effects of
the acceptance, the spin directions are reversed periodically about
every week with polarisation measurements before and after reversal.  The
three-cell configuration results in a more balanced acceptance for cells
with opposite polarisation, which leads to a further reduction of
systematic effects.
% Data taking between two consecutive target spin reversals
%corresponds to the so called `period'. For a period with transversely
%polarised target the polarisations in each cell are measured at the
%beginning and the end of the period.
The achieved polarisation, $P_T$,
is about 0.8 for protons (NH$_3$) and 0.5 for deuterons ($^6$LiD) with
relative uncertainties of $3\,\%$ and $5\,\%$, respectively.

The fraction of polarisable material in the target weighted by the
corresponding cross sections is quantified by the dilution factor, $f$,
that depends on the considered reaction. It is calculated using the
measured material composition of the target and the nuclear dependence
of the cross section for the studied reaction. For incoherent exclusive
$\rho^0$ production the dilution factor is typically 0.25 for
the NH$_3$ target and 0.45 for the $^6$LiD target. See
Sec.~\ref{Sec_extract} for more details.

The target is housed in a large superconducting solenoid providing a
field of 2.5\,T along the beam direction.
% with a field uniformity, $\delta B/B$, better than 10$^{-4}$.
From 2002 to 2004 the angular
acceptance was $\pm$70\,mrad at the upstream edge of the target. From 2006 onwards an upgraded target
magnet with a new large-aperture solenoid was used. It provides an angular
acceptance of $\pm$180\,mrad for the upstream target edge resulting in
an increased hadron acceptance. The transverse holding field of up to 0.5\,T, provided by
a dipole coil, is used for adiabatic spin rotation and for measurements
with transverse target polarisation.

The COMPASS spectrometer is designed to reconstruct scattered muons and
produced hadrons in wide momentum and angular ranges. It consists of
two stages, each equipped with a dipole magnet, to measure tracks with
large and small momenta, respectively.  In the high-flux region, in or close to the
beam, tracking is provided by stations of scintillating fibres, silicon
detectors, micromesh gaseous chambers and gas electron multiplier
chambers. Large-angle tracking devices are multiwire proportional
chambers, drift chambers and straw detectors. Muons are identified in
large-area mini drift tubes and drift tubes placed downstream of hadron
absorbers. Each stage of the spectrometer contains an electromagnetic
and a hadron calorimeter. The identification of charged particles is
possible with a RICH detector, although in this analysis we have not
utilised the information from the RICH.

The data recording system is activated by several triggers. For
inclusive triggers, the scattered muon is identified by a coincidence
of signals from trigger hodoscopes. Semi-inclusive triggers select events
with a scattered muon and an energy deposit in a hadron calorimeter
exceeding a given threshold.  Moreover, a pure calorimeter trigger with
a high energy threshold was implemented to extend the acceptance towards
high $Q^2$ and $x_{Bj}$. In order to suppress triggers due to halo
muons, veto counters upstream of the target are used. The COMPASS
trigger system covers a wide range of $Q^2$, from quasi-real
photoproduction to deep inelastic interactions.
  
%A more detailed description of the COMPASS apparatus can be found in Ref.~\cite{compass}

\section{Event sample}
 \label{Sec_sample}

\begin{figure}[htb]
 \begin{center}
\begin{minipage}[t]{0.48\textwidth} 
\begin{center}
 \epsfig{figure=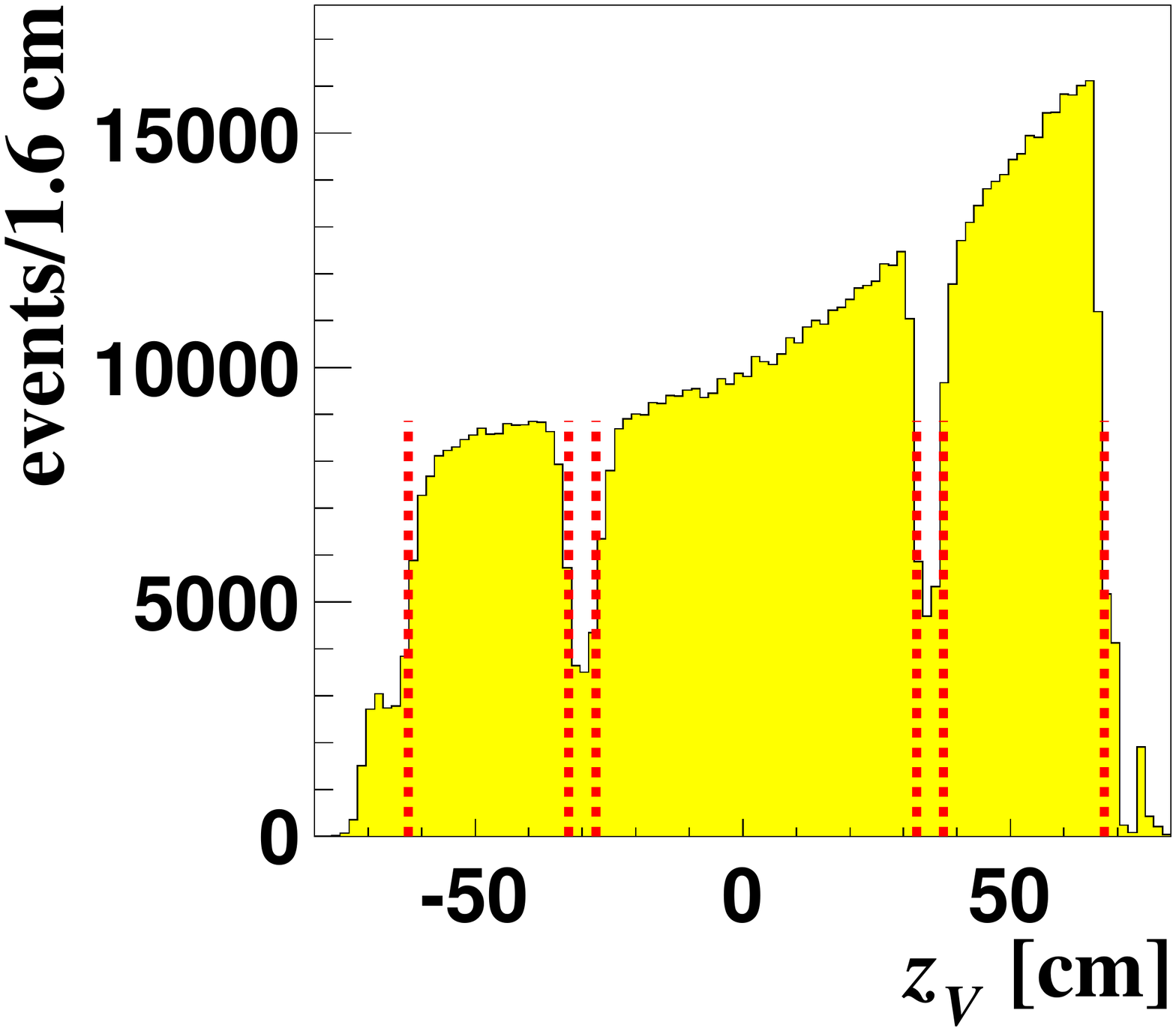,height=0.85\textwidth,clip}
 \end{center} 
  \end{minipage}%
  \begin{minipage}{0.04\textwidth} 
     \hfill %
  \end{minipage}% 
  \begin{minipage}[t]{0.48\textwidth} 
    \begin{center}
 \epsfig{figure=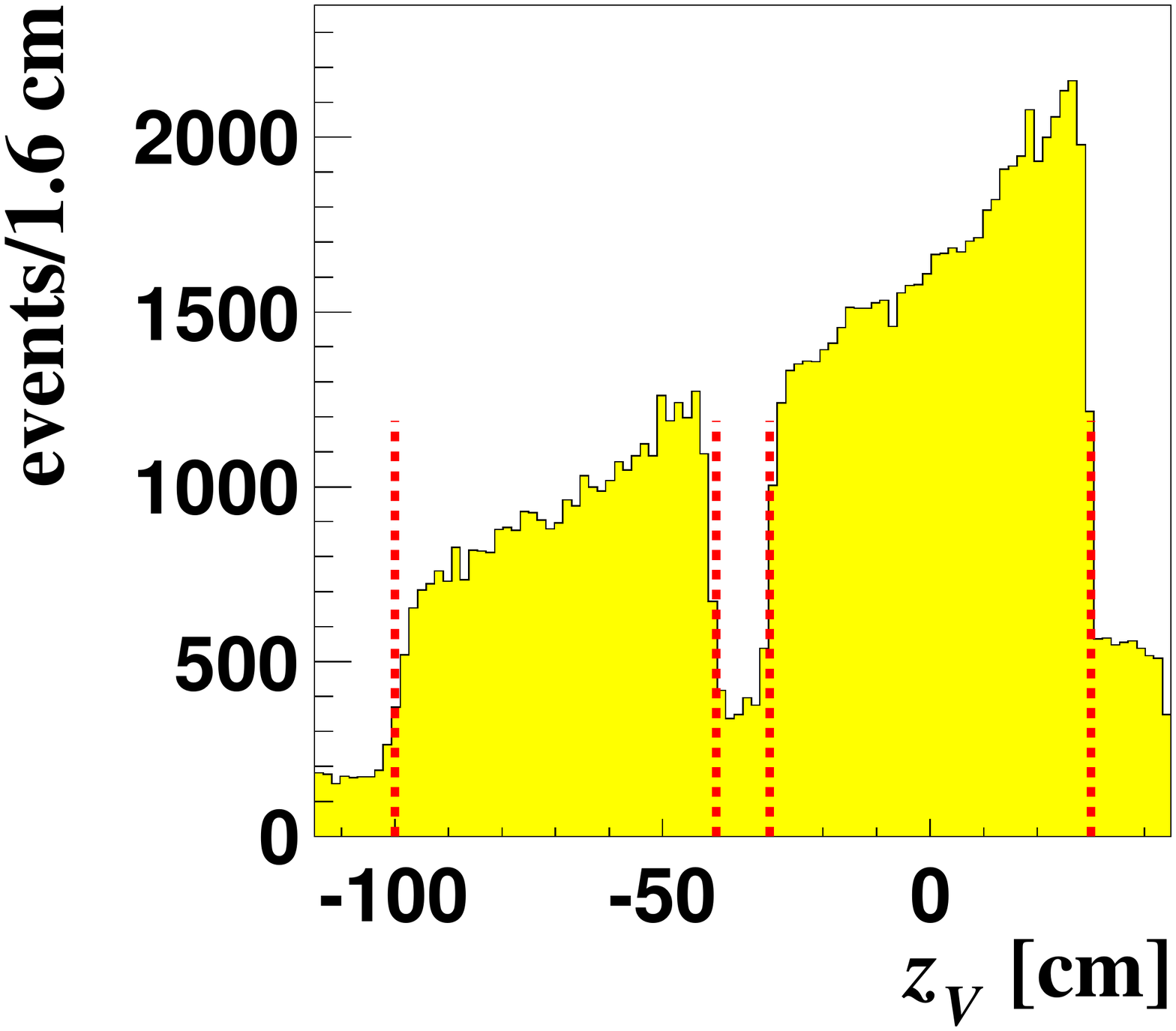,height=0.85\textwidth,clip}
\end{center} 
  \end{minipage}
 \caption{Distributions of the $z$-coordinate of the primary vertex $z_{V}$ for 
NH$_3$  (left) and $^6$LiD data (right).
The vertical lines indicate the applied $z_{V}$ cuts.}
 \label{zvertex}
 \end{center}
\end{figure}

\begin{figure}[htb]
  \begin{center}
  \epsfig{figure=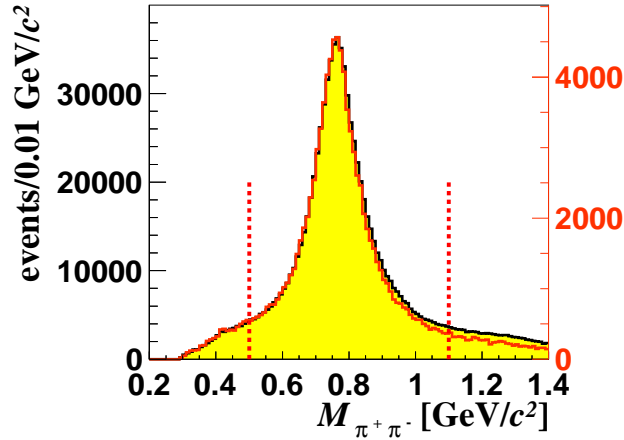,height=7.0cm,clip}
  \caption{Distributions of $M_{\pi \pi}$ for the
 NH$_3$ (black, left scale) and $^6$LiD (red, right scale) data. 
Vertical lines indicate the applied cuts. In order to exclude events originating from
 production of $\phi\/$ mesons decaying into two charged kaons,
 the cut $M_{K K} > 1.04\, {\rm GeV}/c^2$ is applied, where $M_{K K}$ is
 the invariant mass of the two hadron system calculated assuming that
 both hadrons are kaons.}
  \label{masspl}
  \end{center}
 \end{figure}

\begin{figure}[htb]
   \begin{center}
  \epsfig{figure=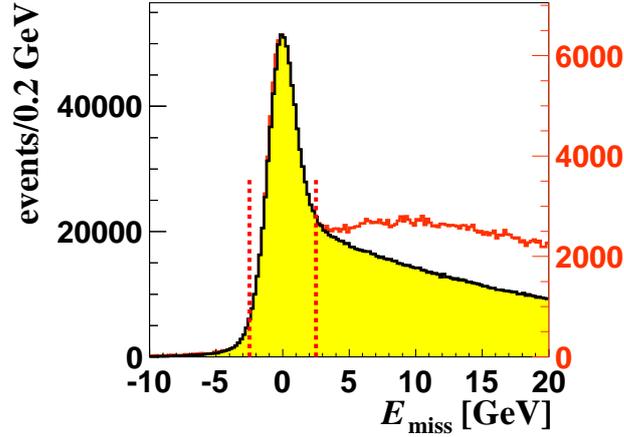,height=7.0cm,clip}
   \caption{Distributions of $E_{\textrm{miss}}$ for the
  NH$_3$ (black, left scale) and $^6$LiD (red, right scale) data. Vertical lines indicate the applied cuts
  to select the signal.}
   \label{emisspl}
   \end{center}
 \end{figure}

\begin{figure}[htb]
 \begin{center}
  \epsfig{figure=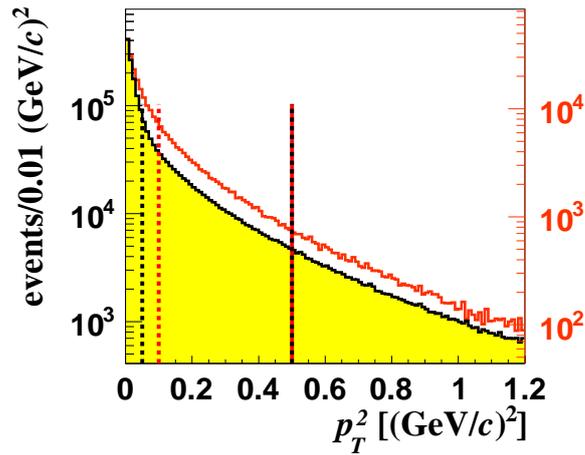,height=7.0cm,clip}
 \caption{Distributions of $p_T^2$ for the
NH$_3$ (black, left scale) and $^6$LiD (red, right scale) data. Vertical lines indicate the applied cuts.
}
 \label{pt2pl}
 \end{center}
\end{figure}

The results presented in this paper are based on the data taken with the
transversely polarised $^6$LiD target in 2003-2004 and with the
transversely polarised NH$_3$ target in 2007 and 2010. The phase space
of the incoming beam is equalised for all target cells using appropriate
cuts on position and angle of beam tracks. An event to be accepted for
further analysis is required to have an incident muon track, a scattered
muon track and exactly two additional tracks of oppositely charged
hadrons, all associated to a vertex in the polarised target
material. Figure~\ref{zvertex} shows the distribution of the reconstructed vertex position $z_{V}$
along the beam axis. In this figure as well as in Figs.~\ref{masspl} to~\ref{pt2pl} the distributions are obtained applying all cuts
except those corresponding to the displayed variable.

In order to obtain a data sample in the deep inelastic scattering
region, the following kinematic cuts are applied: 1\,(GeV/$c$)$^2$
$<$ $Q^2$ $<$ 10\,(GeV/$c$)$^2$, where the upper limit is chosen to
remove the region of $Q^2$ where the fraction of non-exclusive
background is large; $0.1 < y < 0.9$, in order to remove events with
large radiative corrections (large $y$) or poorly reconstructed
kinematics (low $y$). The second cut removes also events from the region
of hadron resonances at small values of $W$.  A small residual number of
such events is removed by requiring $W$ to be larger than 5\,GeV/$c^2$.

As RICH information is not used in this analysis, the charged
pion mass hypothesis is assigned to each hadron track.
In order to select events of incoherent exclusive $\rho^0$ production the following 
additional cuts are applied, which will be justified below:
% on the invariant mass of two pions $M_{\pi \pi}$,
\begin{equation}
\label{masscut}
0.5\,{\rm GeV}/c^2 < M_{\pi \pi} < 1.1\,{\rm GeV}/c^2 ,
\end{equation}
\mbox{~~~}
%on the missing energy $E_{\textrm{miss}}$
\begin{equation}
\label{Emisscut}
-2.5\, {\rm GeV} < E_{\textrm{miss}} < 2.5 \, {\rm GeV} ,
\end{equation}
\mbox{~~~}
%on the transverse momentum squared $p_T^2$  targets
\begin{equation}
\label{erho}
E_{\rho ^0} > 15\, {\rm GeV} ,
\end{equation}
%\begin{equation*}
\mbox{and}
%\end{equation*}
\begin{equation}
\phantom{00}0.1 \, ({\rm GeV}/{\it c})^2 < p_T^2 < 0.5 \, ({\rm GeV}/{\it c})^2 ~~~~~\mbox{ for $^6$LiD} 
\label{ptrange_d}
\end{equation}
%\begin{equation*}
\mbox{or}
%\end{equation*}
\begin{equation}
\phantom{00}0.05 \, ({\rm GeV}/{\it c})^2 < p_T^2 < 0.5 \, ({\rm GeV}/{\it c})^2 ~~~~~\mbox{ for NH$_3$}.
\label{ptrange_p}
\end{equation}
These cuts allow us to minimise the effects of various types of
backgrounds such as: (i) semi-inclusive deep-inelastic (SIDIS) production of a $\rho
^0$ meson or $\pi^+\pi^-$ pair, (ii) $\rho ^0$ production with
diffractive dissociation of the target nucleon, (iii) exclusive
non-resonant $\pi^+\pi^-$ pair production, (iv) coherent exclusive
$\rho ^0$ (or non-resonant $\pi^+\pi^-$ pair) production on a target nucleus.

Figure~\ref{masspl} shows the distributions of $M_{\pi \pi}$ for the NH$_3$ and $^6$LiD
targets. A clear peak of the $\rho ^0$ resonance is visible
on top of a background arising from (i) and (iii). The selection 
on $M_{\pi\pi}$ (Eq.~\eqref{masscut}) is optimised to minimise the effect of exclusive
non-resonant $\pi^+\pi^-$ pair production (iii) that will be
discussed in Sec.~\ref{Background}.

Because slow particles are not detected, exclusive events are
selected by the cut on missing energy given by
Eq.~\eqref{Emisscut}. The selected range is referred to as `signal region' in the following.
In the $E_{\textrm{miss}}$ distributions presented in Fig.~\ref{emisspl}
the peak at $E_{\textrm{miss}} \approx 0$ is the signal of exclusive
$\rho ^0$ production. The width (rms) of the peak, $\sigma \approx 1.25
\,\rm{GeV}$, is due to spectrometer resolution which motivates the cut on $E_{\textrm{miss}}$ (Eq.~\eqref{Emisscut}). Non-exclusive events (i) and (ii),
where in addition to the recoil nucleon other undetected hadrons are
produced, appear at $E_{\textrm{miss}}$ above about zero. However, due to the finite
resolution they cannot be resolved from the exclusive peak. The observed difference of $E_{\textrm{miss}}$
distribution shapes between the two samples at large $E_{\textrm{miss}}$
is due to the increase of the angular acceptance of the COMPASS setup mentioned in
Sec.~\ref{Sec_exper}.

The $p_T^2$ distributions are shown in Fig.~\ref{pt2pl}. 
%We choose to
%use $p_T^2$ rather than $|t|$; in the COMPASS kinematic region the two
%variables are approximately equal\footnote{and another footnote} but the
%former is determined with better precision, by a factor of 2 to 3.
We choose to use $p_T^2$ rather than $t$ or
$t'= |t| -t_{0}$, where $t_{0}$ is the minimal kinematically allowed $|t|$, 
because in the COMPASS kinematic 
region $p_T$ is determined with better precision by a factor of
2 to 5. In addition, the $t'$ distribution is distorted because $t_0$, which
depends on $W$, $Q^2$, $M_{\pi\pi}$ and $M_X^2$, is poorly
determined for non-exclusive background events~\cite{NMC-rho-1}. 
The shown $p_T^2$ distributions indicate at small $p_T^2$ values
contributions from coherent $\rho ^0$ production on target nuclei.
Coherent events are suppressed by applying the lower cuts given by
Eqs~(\ref{ptrange_d},\ref{ptrange_p}). A study of $p_T^2$ distributions
shows that in addition to exclusive coherent and incoherent $\rho ^0$
production a third component, originating from non-exclusive background (i),
is also present and its contribution increases with $p_T^2$, thus requiring
also an upper cut.  Therefore,
in order to select the sample of events from incoherent exclusive $\rho
^0$ production, the afore-mentioned $p_T^2$ cuts were applied, which are
indicated by vertical lines in Fig.~\ref{pt2pl}.
% Semi-exclusive background (i) is further reduced
%with the upper cut on $p_T^2$, Eqs~(\ref{ptrange_d},\ref{ptrange_p}),
%and the cut on $E_{\rho ^0}$ (Eq.~\eqref{erho}).

After all selections the final samples for incoherent exclusive $\rho ^0$
production consist of about 797000 for the NH$_3$ target and 97000 events for $^6$LiD
target. The mean values of the kinematic variables
$Q^2$, $x_{Bj}$, $y$, $W$ and $p^2_T$ are given in Table~\ref{tab:kin_mean}.

\begin{table}[!ht]
\begin{center}
\caption{Mean values of the kinematic variables for proton and deuteron data.}
\label{tab:kin_mean}
\vspace{0.35cm}
\begin{tabular}{c c c c c c }
\toprule
& $\langle Q^2 \rangle$ $(\rm{GeV}/\mathit{c})^2$ & $\langle x_{\mathit{Bj}} \rangle$ & $\langle y\rangle$ & $\langle W \rangle$ $(\rm{GeV}/\mathit{c}^{2})$ & $\langle p^2_T \rangle$ $(\rm{GeV}/\mathit{c}^2)$\\
\midrule
proton data   & 2.2 & 0.039 & 0.24 & 8.1 & 0.18 \\
deuteron data & 2.0 & 0.032 & 0.27 & 8.6 & 0.23 \\
\bottomrule
%\multicolumn{1}{c}{\rule{0.09\textwidth}{0px}} &
%\multicolumn{1}{c}{\rule{0.09\textwidth}{0px}} &
%\multicolumn{1}{c}{\rule{0.09\textwidth}{0px}} &
%\multicolumn{1}{c}{\rule{0.09\textwidth}{0px}} &
%\multicolumn{1}{c}{\rule{0.09\textwidth}{0px}} &
%\multicolumn{1}{c}{\rule{0.09\textwidth}{0px}}
\end{tabular}
\end{center}
\end{table}

\section{Background to exclusive $\rho ^0$ production}
\label{Background}

The most important background contributions introduced in the previous
section are discussed here in more detail.

({\bf i}) The SIDIS contribution constitutes the largest source of
background to the exclusive sample. It is estimated using Monte Carlo
(MC) samples generated by the LEPTO generator with the COMPASS tuning
of JETSET parameters~\cite{highptpairs}. The detector response is
simulated using the description of either the NH$_3$  or the $^6$LiD 
set-up for transverse target polarisation. Simulated data are subject to
the same selection criteria as real data and analysed in the same bins
of kinematic variables. Comparing the $E_\textrm{miss}$ distributions of
real and simulated data reveals insufficient agreement for
$E_{\textrm{miss}} > 7$\,GeV, where only SIDIS background
contributes. The situation is improved considerably by weighting the
$h^+ h^-$ MC data in every $E_\textrm{miss}$ bin $i$, by the ratio of
numbers of like-sign events %($h^+ h^+$ and $h^- h^-$) from real and MC
from real and MC data:
\begin{equation}
\label{weight}
w_i^{\rm like} = \frac{N^{h^+ h^+}_{i,real}+N^{h^- h^-}_{i,real}}{N^{h^+ h^+}_{i,MC}+N^{h^- h^-}_{i,MC}},
\end{equation}
independent of other kinematic variables.  As like-sign data contain
only background, the weighting can be applied over the full
$E_\textrm{miss}$ range. This weighting procedure relies on the
assumption that weights obtained from like-sign data are applicable to
unlike-sign data. This assumption is supported by the observation that
$w_i^{\rm like} \simeq w_i^{\rm unlike} \equiv {N^{h^+
h^-}_{i,real}}/{N^{h^+ h^-}_{i,MC}}$ holds at large $E_{miss}$, despite
of different shapes of $E_\textrm{miss}$ distributions. The shape of the
resulting weighted $E_{miss}$ distribution for unlike-sign MC data is
parameterised for each individual target cell in every bin of $Q^2$,
$x_{Bj}$ or $p_T^2$. As the acceptance does not show a $\phi-\phi_S$
dependence, the MC data is not binned in this variable.

\begin{figure}[htb]
 \begin{center}
\begin{minipage}[t]{0.48\textwidth} 
\begin{center}
 \epsfig{figure=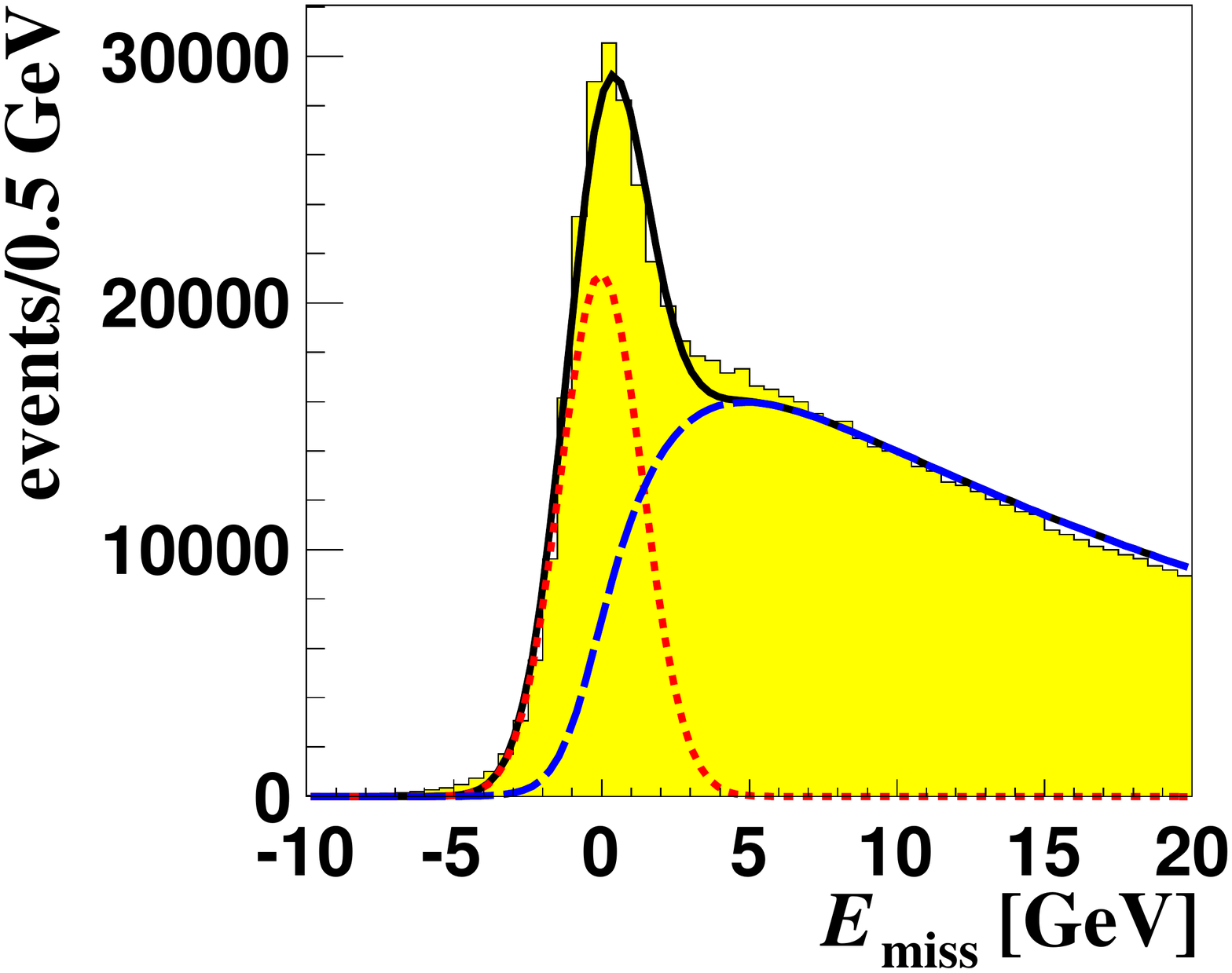,height=0.85\textwidth,clip}
 \end{center} 
  \end{minipage}% Dies Prozent ist wichtig! (kein horiz. Abst. zw. minipages) 
  \begin{minipage}{0.04\textwidth} 
     \hfill % Damit die getrennte Beschriftung auch Abstand hat 
  \end{minipage}% 
  \begin{minipage}[t]{0.48\textwidth} 
    \begin{center}
 \epsfig{figure=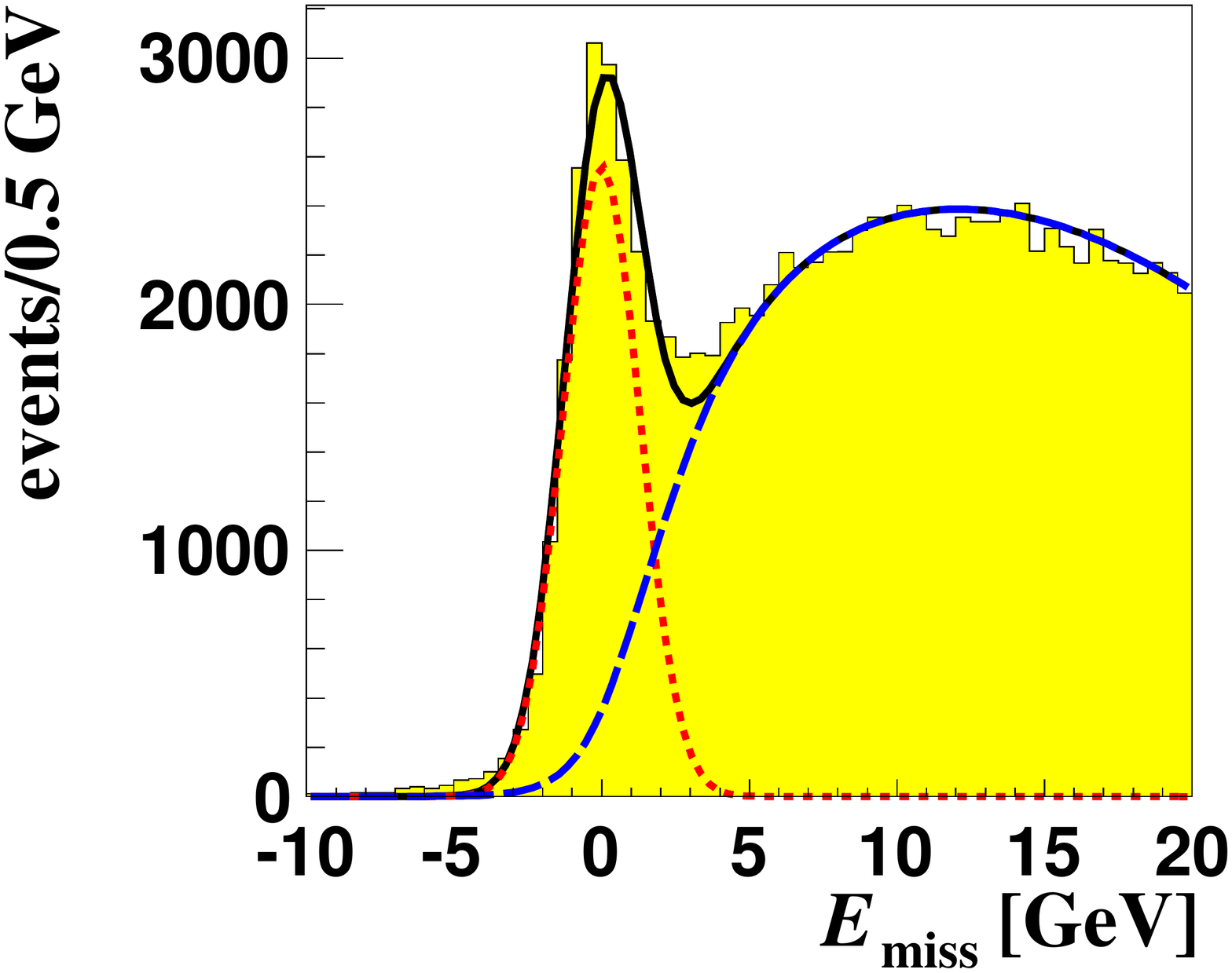,height=0.85\textwidth,clip}
\end{center} 
  \end{minipage}
 \caption{The $E_\textrm{miss}$ distributions in the range
$2.4$\,(GeV/$c)^2 < Q^2 \leq 10.0$\,(GeV/$c)^2$, together with signal plus
background fits (solid curves) for the NH$_3$ (left) and $^6$LiD (right) 
samples. The dotted and dashed curves represent the signal and background 
contributions, respectively.}  
 \label{fig:kin_var_3}
 \end{center}
\end{figure}

For the determination of the asymmetries as described in
Sec.~\ref{Sec_extract}, the NH$_3$ and $^6$LiD real data are binned in
the same way in $Q^2$, $x_{Bj}$ or $p_T^2$ per target cell, and also in
\mbox{$\phi-\phi_S$} and according to the target spin orientation. Here,
a binning in \mbox{$\phi-\phi_S$} is preferred over a simultaneous
binning in $\phi$ and $\phi_S$ due to lack of statistics. In every such
bin, the $E_{\textrm{miss}}$ distribution is fitted using a Gaussian for
the signal of exclusive events in conjunction with the above explained
fixed shape for the SIDIS background, with free normalisation. The data
are corrected for this background on a bin-by-bin basis. As an example,
Fig.~\ref{fig:kin_var_3} illustrates the two-component fit for the $Q^2$ bin with the
largest background contribution.
From these fits, the fractions $f_{\textrm{sidis}}$ of SIDIS events in
the signal region (see Eq.~\eqref{Emisscut}) are found to vary between
0.05 and 0.4, depending on kinematics and target set-up. On average,
$f_{\textrm{sidis}}$ equals 0.18 for deuteron and 0.22 for proton data.

%as described in Sec.~\ref{Sec_extract}.
%Due to a non-negligible fraction of the SIDIS background
%the results have to be corrected by subtracting estimated numbers of
%background events, as described in Section~\ref{Sec_extract}.
%ABOVE SENTENCE PRESENTLY OUTCOMMENTED, AS I CAN'T UNDERSTAND THE MEANING.

({\bf ii}) Diffractive dissociation of the target nucleon into several
particles is another type of background.  %a baryon and additional ,
%$\gamma ^*N \rightarrow \rho ^0N^*$, in which a target %nucleon is
%excited to a baryonic resonance $N^*$ (or $\Delta$) that %subsequently
%decays into a nucleon and one or several mesons.  
Without a recoil
detector, such events cannot be resolved from exclusive events by
requirements on missing energy unless the mass $M_{X}$ of the recoiling
system exceeds 2.4\,GeV/$c^2$. The high-$M_{X}$ tail of diffractive
dissociation events can be seen in Fig.~\ref{fig:kin_var_3} next to the
exclusive peak as a small enhancement over the SIDIS background. 
Diffractive dissociation background is examined using a MC event generator called 
HEPGEN~\cite{hepgen}. This generator is dedicated to studies of hard exclusive
single photon or meson production processes in the COMPASS kinematic domain.
In addition, it allows also to generate single photon or meson production accompanied by 
diffractive dissociation of the nucleon. 
Using both exclusive and nucleon-dissociative $\rho ^0$ events generated by HEPGEN, 
%Using
%nucleon-dissociative events generated by HEPGEN~\cite{hepgen}\footnote{this sentence will be changed in a smart way},
which are reconstructed and selected as the real data, 
the contribution from low-mass diffractive dissociation 
of the nucleon is found to be $\approx 14\,\%$ of the incoherent exclusive
$\rho^0$ signal. No attempt is made to remove this type of background,
motivated by HERA results on $\rho ^0$ production where for
unpolarised protons the angular distributions of proton-dissociative
events were found consistent with those of exclusive events~\cite{zeus1,
zeus2, h1}. Using the $E_{\textrm{miss}}$ shape from HEPGEN,
three-component fits to the experimental $E_{\textrm{miss}}$
distributions show negligible impact of nucleon-dissociative events onto
the determination of $f_{\textrm{sidis}}$.

\begin{figure}[htb]
 \begin{center}
 \epsfig{figure=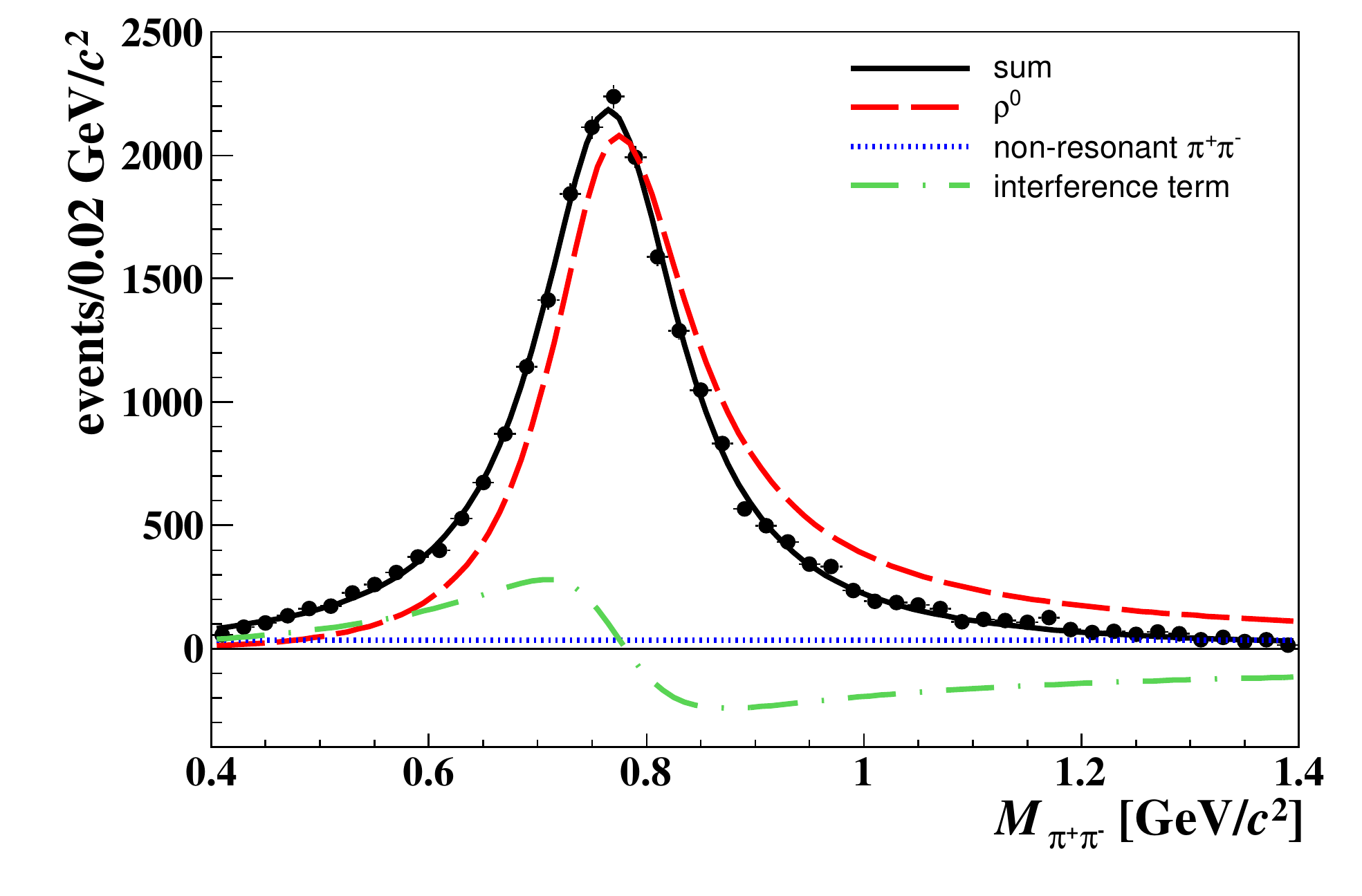,height=5.5cm,clip}
 \caption{The $M_{\pi^{+}\pi^{-}}$ distribution with a fit using the S\"{o}ding
parameterisation is shown for NH$_3$ data in the selected kinematic 
range $1$\,(GeV$/c)^2 < Q^{2} < 1.2$\,(GeV$/c)^2$, 
using the restrictive cut 
$-2.5\,\textrm{GeV} < E_{\textrm{miss}} < 0\,\textrm{GeV}$ to suppress the 
semi-inclusive background. The thick solid line represents the result of the fit, 
while the dashed, dotted and dashed-dotted ones
 represent contributions from resonant $\rho^0$ production, non-resonant
$\pi^+\pi^-$ production and the interference term respectively.}
 \label{massfit}
 \end{center}
\end{figure}

({\bf iii}) The background contribution due to exclusive non-resonant
$\pi^+ \pi ^-$ production, in particular the impact of its interference with resonant $\rho ^0
\rightarrow \pi ^+ \pi ^-$ production, is studied in bins of $Q^{2}$,
$x_{Bj}$ or $p_{T}^{2}$, while suppressing the SIDIS background by a
restrictive cut on missing energy, $-2.5\,\textrm{GeV} <
E_{\textrm{miss}} < 0\,\textrm{GeV}$.  As an example, the $M_{\pi ^+\pi
^-}$ distribution is shown in Fig.~\ref{massfit} for a selected $Q^2$
range, $1$\,(GeV$/c)^2 < Q^{2} < 1.2$\,(GeV$/c)^2$.  The modification of the $\rho^0$ 
relativistic Breit-Wigner shape in the presence of non-resonant $\pi
^+ \pi ^-$ events, which is observed for the distribution, is taken into
account by applying either the S\"{o}ding~\cite{soding} or the
Ross-Stodolsky~\cite{RossStod} approach.  The result of the fit to the
data using the S\"{o}ding parameterisation is shown in the figure, where
contributions from the $\rho ^0$ resonance, non-resonant $\pi ^+ \pi ^-$
pair production and the interference term are displayed.  Different
ranges of the invariant mass $M_{\pi ^+\pi ^-}$ were examined in order
to minimise the impact of non-resonant pair production and the
interference term. The determined range is given by Eq.~\eqref{masscut}, leading to
an overall effect of less than 2\,\% in any kinematic bin of $Q^2$,
$x_{Bj}$ or $p_T^2$.  Exclusive resonant $\rho ^0$ and exclusive non-resonant pair
contributions are not distinguished in the following.
%, as commonly done
%in previous theoretical and experimental publications (see {\it e.g.} 
%Ref.~\cite{hermes3} and references therein).

({\bf iv}) Coherent exclusive $\rho ^0$ production on various nuclei of
the target constitutes additional background. 
Its magnitude is estimated from the analysis of the shape of $p_T^2$ distributions.
In the kinematic region
defined by Eqs~\eqref{masscut}-\eqref{ptrange_p}, it amounts to 
$\simeq 12\,\%$ for NH$_3$ and $\simeq 8\,\%$ for $^6$LiD.
%using fits of the sum
%of three exponential functions to the $p_T^2$ distributions. In
%addition, for the deuteron target the $p_T^2$ distributions were
%calculated using the Glauber model inspired calculations, which przeus2,h1}.ovide
%more realistic description of coherent and incoherent contributions than
%three-exponential fits.  The estimated contribution of coherent events
%in our standard sample is
%This contribution is neglected, which is justified by the observation
No correction is applied for this residual background, which is justified by the observation
that in this region the asymmetry \AUT is consistent within statistical uncertainty
with that for
events from the small-$p_T^2$ region. The latter is defined by $p_T^2 < 0.05\,
(\rm{GeV}/{c})^2$ for the NH$_3$ target and $p_T^2 < 0.10\,
(\rm{GeV}/{c})^2$ for the $^6$LiD target, where coherent production
dominates.

%In order to maximise the "purity" of theEq.~\eqref{masscut}
%$\rho^0$ sample we studied the ratio of the number of $\rho^0$ events
%integrated between $M_{\pi\pi}^{\textrm{min}}$ and
%$M_{\pi\pi}^{\textrm{max}}$ with respect to the number of all events
%($\rho^0$, non-resonant $\pi^{+}\pi^{-}$ and interference) integrated
%between $M_{\pi\pi}^{\textrm{min}}$ and
%$M_{\pi\pi}^{\textrm{max}}$. With such a definition the "purity" can be
%either smaller or larger than the unity, because the overall
%contribution of the interference term may be either positive or
%negative, depending on the selected range of $M_{\pi\pi}$.
%With an optimal choice of $M_{\pi\pi}$ range, which is defined by
%Eq.~\eqref{masscut}, the high purity, close to 100\,\% is achived. In
%terms of the S\"{o}ding approach, it may be explained, for this
%particular choice of $M_{\pi\pi}$ range, by an approximate compensation
%of the positive and negative interference term contributions
%respectively at masses below and above the nominal $\rho ^0$ mass.

\section{Extraction of the asymmetry}
 \label{Sec_extract}

The asymmetry is determined in bins of $Q^2$, $x_{Bj}$ or $p_T^2$, while
integrating over the two remaining variables. For brevity the dependence
on these variables is omitted in the following.  The number of exclusive
events, after subtraction of
SIDIS background, can be expressed as a function of the angle $\phi-\phi_S$ in the following way
\begin{equation}
\label{eq:method_1} N( \phi-\phi_S ) = F\,n\,a\, \sigma_0 \left( 1 \pm  f\,|P_T|\,
A_{UT}^{\sin(\phi-\phi_S)} \sin ( \phi-\phi_S ) \right),
\end{equation}
where $F$ is the muon flux, $n$ the number of target nucleons,
$a(\phi-\phi_S)$ the product of acceptance and efficiency of the
apparatus, $\sigma _0$ the spin-averaged cross section, $f$ the dilution
factor, $\pm |P_T|$ the target polarisation, and the asymmetry \AUT is
defined by Eq.~\eqref{eq:access5}. In this analysis, the asymmetry is
extracted from a direct fit of the number of events in bins of $\phi -
\phi_S$. For each of the two target cells (${ncell}=1,2)$ and
polarisation states $(+,-)$,
\begin{equation}
\label{eq:method_2}
%N^{\pm}_{j,{ncell}}=c^{\pm}_{j,{ncell}} (1 \pm A \sin(\phi - \phi_S))
%N^{\pm}_{j,{ncell}} = \frac {1}{\Delta \phi} \int_{\phi_j}^{\phi_j+\Delta \phi} \textrm{d}\phi\,c^\pm_{j,ncell}(1\pm A\sin(\phi))
N^{\pm}_{j,{ncell}} = \frac {1}{\Delta } \int_{\phi_{c,j}-\Delta /2}^{\phi_{c,j}+\Delta /2} \textrm{d}\phi'\,c^\pm_{j,ncell}(1\pm A\sin(\phi'))
\end{equation}
gives the number of events in bin $j$ of $\phi - \phi_S$, where
$j=1$ to $m$, and $m=12$ the number of bins. 
Here $\phi' = \phi - \phi_S$, the central value of bin $j$ is denoted 
by $\phi_{c,j}$ and $\Delta $ is the bin width. 
In the case of NH$_3$ data,
events from upstream and downstream cells of the three-cell target are
combined. This leads to a system of $2\times 2\times m$ non-linear
equations.  The `normalised acceptance' $c^{\pm}_{j,{ncell}}$ is the
product of spin-averaged cross section, muon flux, number of target
nucleons, acceptance and efficiency of the apparatus. The dependence on
target dilution factor $f$ and target polarisation $P_T$ is absorbed
into the `raw asymmetry' $A$ that is directly fitted to the data.

A one-dimensional binned maximum likelihood method is used to solve the
system of 4$m$ equations. Here the likelihood is constructed from
Gaussian distributions in order to account for the non-Poissonian nature
of the background subtracted data. In order to reduce the number of
unknowns, the reasonable assumption is made that possible changes of
acceptances in target cell before and after target polarisation
reversal are the same for every bin $j$ and can be described by a common constant $C$:
\begin{equation}
\label{eq:RA}
C=\frac{c_{j,1}^+ c_{j,2}^+}{c_{j,1}^- c_{j,2}^-}.
\end{equation} 
 Using this constraint, 
one can determine the 3$m$ independent
normalised acceptances $c^{\pm}_{j,{ncell}}$, the constant $C$ and the raw asymmetry $A$.
% The correction for the SIDIS background is performed individually for
%  each kinematic bin in $x_{Bj}$, $Q^2$ and $p^2_T$, each cell and
%  polarisation state, and each $\phi - \phi_S$ bin. The signal plus
%  background fits were performed as described in Sec.~\ref{Background}.
%  In order to be less sensitive to statistical fluctuations in the Monte
%  Carlo sample, the used background shapes were determined from
%  $\phi-\phi_S$ integrated distributions, and only the normalisation of
%  the shape was fitted to data independently in each $\phi-\phi_S$
%  bin. The estimated number of SIDIS background events in the signal
%  region was then subtracted from the number of observed events on a
%  bin-by-bin basis.  In this approach one relies on the assumption that
%  the background asymmetry of the $\phi-\phi_S$ distribution does not
%  depend on $E_\textrm{miss}$. It is supported by studies of the asymmetry
%  as a function of $E_\textrm{miss}$ in the range $-2.5 < E_\textrm{miss}
%  < 20$\,GeV where no such dependence was found within statistical
%  precision. 
%The results of the fits to the background-subtracted numbers
% of events are corrected to account for the finite bin size.

\begin{figure}[htb]
 \begin{center}
 \epsfig{figure=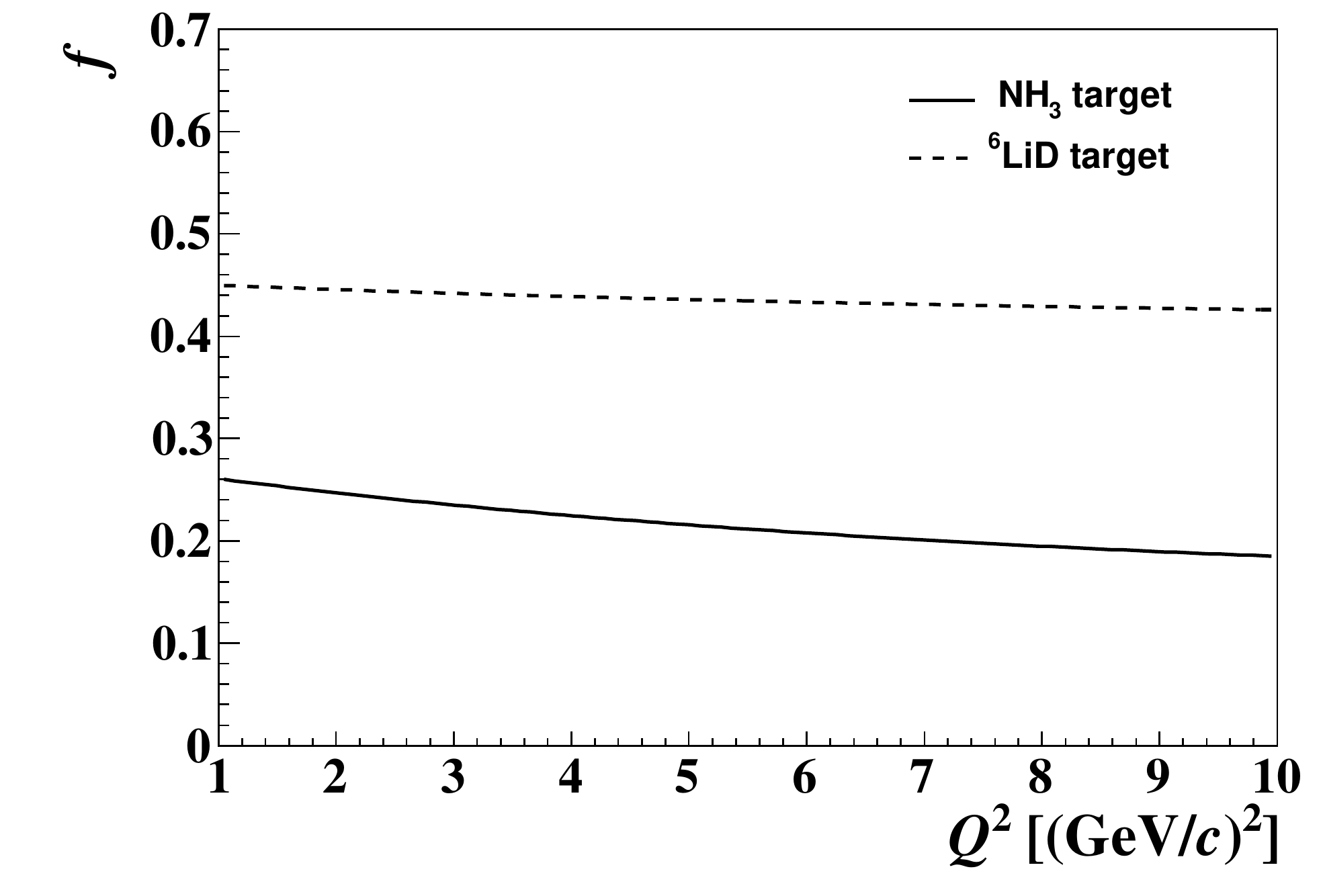,height=5.5cm,clip}
 \caption{Dilution factor $f$ for exclusive $\rho^0$ production as a function of $Q^2$. The
solid (dashed) line represents the dilution factor for the NH$_3$
($^6$LiD) target.}
 \label{dilfac}
 \end{center}
\end{figure}

In every bin $x_{Bj}$, $Q^2$ or $p_T^2$, the asymmetry \AUT is calculated 
as \AUT = $A/\langle f\: |P_T| \rangle$ using the raw asymmetry $A$ obtained from the fit.
%The evaluation of  the physics asymmetry \AUT  requires  
%mean values of $P_T$ and  $f$. %$\langle f |P_T| \rangle$
The dilution factor $f$ is calculated on an event-by-event basis using
the measured contributions of various atomic elements in the target and
a parameterisation of the nuclear dependence of the spin-independent
cross section for the studied reaction, $\mu  N \rightarrow \mu 
\rho ^0 N$, as explained in Ref.~\cite{rholong}. The ratios of this
cross section per nucleon for a given nucleus to the cross section on
the proton or deuteron are parameterised~\cite{ATrip} over a wide $Q^2$-range, using
measurements on various nuclear targets. %performed by the NMC, E665 and
%photoproduction experiments.  
No dependence of nuclear effects on $y$ or
$\nu $ is assumed, motivated by NMC results on exclusive $\rho ^0$
production \cite{NMC-rho-2} in a kinematic range similar to that of COMPASS. 
The $Q^2$ dependence of the dilution factor for incoherent exclusive $\rho ^0$ production
is shown in Fig.~\ref{dilfac}. As can be seen, the values
of the dilution factor $f$ for
the NH$_3$ target vary from 0.27 at $Q^2 = 1 \,(\rm{GeV}/{\it c})^2$ to 0.18 at
$Q^2 = 10 \,(\rm{GeV}/{\it c})^2$, and correspondingly from 0.45 to 0.42 for 
the $^6$LiD target\footnote{Our estimates of $f$ for the $^6$LiD target are by 9
- 19\,\% higher than those of Ref.~\cite{rholong}. The difference, which
is due to an improvement of the treatment of the $^6$Li nucleus, is
significantly smaller than the total systematic uncertainty quoted in
Ref.~\cite{rholong}.}.
Radiative corrections are neglected in the
present analysis, in particular in the calculation of $f\/$. They are
expected to be small, mainly because of the exclusivity
cuts (see Sec.~\ref{Sec_sample}) that largely suppress the otherwise
dominant external photon radiation~\cite{akushevich}. 
%Radiative corrections are neglected in the
%present analysis, in particular in the calculation of $f\/$. They are
%expected to be small, mainly because of the exclusivity
%cuts (see Sec.~\ref{Sec_sample}) that largely suppress the otherwise
%dominant external photon radiation~\cite{akushevich}.

The correction to the proton asymmetry $A_{UT,p}^{\rm{sin}(\phi-\phi_S)}$ due to the polarisation
of $^{14}$N nuclei in the ammonia target was estimated following the approach
of Refs~\cite{rondon,sidis2012}. The correction is proportional to the measured asymmetry for the deuteron and approximately given by:
\begin{equation}
\label{eq:corrN14}
\Delta A_{UT,p}^{\rm{sin}(\phi-\phi_S)} = \frac{1}{3}\cdot(-\frac{1}{3})\cdot
\frac{1}{6}\cdot \frac{\sigma _d}{\sigma_p}\cdot A_{UT,d}^{\rm{sin}(\phi-\phi_S)}\: .
\end{equation}
The factors account for the fraction of polarisable nitrogen nuclei
in ammonia, the alignment of proton spin vs. $^{14}$N spin, the ratio
of $^{14}$N to $^1$H polarisations and the ratio of  
cross sections, $\sigma _d$
and $\sigma _p$, for exclusive $\rho ^0$ production by scattering muons off
unpolarised deuteron and proton targets, respectively.
The estimated corrections are very small, typically about 0.1\,\%,
and are neglected in the following.

\section{Systematic uncertainties}
\label{Systematics}

In this section we describe tests performed to examine various sources
of possible systematic uncertainties, namely:
 (a) bias of the applied
estimator of the asymmetry, (b) data stability, (c) false asymmetries, (d) sensitivity to the method of background
subtraction, (e) sensitivity to the Monte Carlo description used for the
parameterisation of the background shape, (f) compatibility of results
after background subtraction, and (g) uncertainties of target dilution
factor and target polarisation value.  As tests (a)--(c) are not sensitive
to the background, they are performed using background non-corrected
data.  Possible systematic effects related to background subtraction are
subject of tests (d)--(f).

{\bf (a)} In this analysis a one-dimensional binned maximum likelihood
method is used (see Sec.~\ref{Sec_extract}) rather than an extended unbinned
maximum likelihood method~\cite{euml-1,euml-2} because the latter one depends
strongly on the quality of the Monte Carlo description for the SIDIS
background in the kinematic region of our data. At present none of the
existing generators satisfactorily describes the SIDIS background. The bin-by-bin comparison of the background non-corrected results for each target indicates good agreement between
both estimators. However, for each of the three binnings used, 
in $x_{Bj}$, $Q^2$ and $p_T^2$, the mean
asymmetry values for the deuteron from the binned method is observed to be
slightly smaller than that
from the unbinned method, by several percent of the statistical uncertainty. 
Additionally, systematic differences of
similar size are seen in the deuteron data
between the mean values of the asymmetries evaluated in the three
binnings. Both these effects result in 
a systematic uncertainty of $\approx 0.10\,\sigma^{stat}$. As systematic 
uncertainties (a)--(f) are evaluated using data, here and in the 
following a systematic uncertainty is expressed in terms of $\sigma^{stat}$,
which is the statistical error of the background
corrected asymmetry measured in a given kinematic bin.
For the proton data
one observes good agreement between different estimators also at the level of mean
asymmetries. The distribution of differences between mean asymmetries obtained from
two estimators is centred at zero and its RMS value is used to estimate a systematic
uncertainty of $\approx 0.12\,\sigma^{stat}$.

{\bf (b)} Subsets of data, each containing two periods consecutive in
time with opposite target polarisations, are compared to test the
stability of data taking. For each of the 18 (3) subsets formed from
proton (deuteron) data, the asymmetry is determined in every kinematic
bin of $x_{Bj}$, $Q^2$ and $p_T^2$.  All asymmetries are found to be
compatible within statistical uncertainties.

{\bf (c)} In order to investigate possible false asymmetries, the target
is artificially divided into four cells of 30\,cm length each,
distributed contiguously along the beam direction. This allows the evaluation of two independent false asymmetries using cells with the same spin orientation.
 Determining these false asymmetries in the exclusive
 region leads to statistical fluctuations similar to those of the physics
 asymmetry. In order to increase the statistical significance, the false
 asymmetries are studied in an extended range, $-10\:$GeV$ < E_\textrm{miss} <
 20$\,GeV.
% %The deviations of $A_f^+$ and $A_f^-$ from
% %zero are calculated as weighted mean deviation, $\Delta _{A_f^{\pm}}$,
% %measured in terms of its statistical uncertainty, $\sigma_{\Delta}^{stat}$,
% In order to quantify deviations of the false asymmetries from zero the
% weighted means, $\Delta _{A_f^{\pm}}$, measured in terms of their
% statistical errors, $\sigma_{\Delta}^{stat}$, is calculated:
% \begin{equation}
% \label{eq:sys_err_add}
% \frac {\Delta_{A_{f}^{\pm}}}{\sigma_{\Delta}^{stat}} = \frac{\sum_{i} \frac{|A_{f,i}^{\pm}|}{\sigma_i} \frac{1}{\sigma_i^2}}{\sum_{i}  \frac{1}{\sigma_i^2}}.
% \end{equation}
% Here the sum runs over 12 (13) kinematic bins and $\sigma_i$ denotes the statistical error of $A_{f,i}^\pm$. 
The resulting false asymmetries are found to be consistent with zero within statistical uncertainties. They are used to estimate upper bounds
for the corresponding systematic uncertainty, namely 
$0.15\, \sigma^{stat}$ ($0.49\, \sigma^{stat}$) for the proton
(deuteron) data. 
%Here $\sigma^{stat}$ is the statistical error of the background
%corrected asymmetry measured in a given kinematic bin.

{\bf (d)} In order to estimate the sensitivity of the extracted
azimuthal asymmetry to the method of background subtraction, an
alternative method is applied using measured $\phi-\phi_S$ distributions
of the SIDIS background. Their shapes are determined in the range $7\,{\rm GeV} <
E_\textrm{miss} < 20$\,GeV, where exclusive and diffractive-dissociation
events can be neglected against the SIDIS background. They are measured
for each kinematic bin in $x_{Bj}$, $Q^2$, and $p^2_T$ respectively, and
for each target cell and polarisation state.
They are rescaled in such a
way that in the exclusive region (see Eq.~\eqref{Emisscut}) the total
number of events for the rescaled distribution is equal to the number of
background events obtained from the signal plus background fit to the
$E_\textrm{miss}$ spectra.
%The properly rescaled spectra are subtracted from the
The rescaled spectra are subtracted from the corresponding
$\phi-\phi_S$ distributions for the data
and the asymmetries are extracted from the resulting
background-corrected distributions as described in Sec.~\ref{Sec_extract}. 
%Here, similarly as for the 
%default method\footnote{???} described in Sec.~\ref{Sec_extract}, one assumes that 
%the background asymmetry does not depend on $E_\textrm{miss}$. 
In this approach one relies on the
assumption that the background asymmetry does not depend on
$E_\textrm{miss}$. It is supported by studies of the asymmetry as a
function of $E_\textrm{miss}$ in the range $-2.5\,{\rm GeV} < E_\textrm{miss} <
20$\,GeV where no such dependence was found within statistical
precision.
Both the default and alternative methods are used to extract \AUT. A
point-by-point comparison in the three kinematic binnings indicates for
all data sets very good agreement between the two methods, within statistical uncertainties.
%, typically
%within $\pm 0.1 \,\sigma _{stat}$. Only in the largest $x_{Bj}$ and
%$Q^2$ bin for proton data a somewhat larger difference ($\simeq 0.5
%\,\sigma _{stat}$) is observed.\\ ANDRZEJ: where THIS (THESE)
%syst. uncertainties appear in table 3 ??? \\

{\bf (e)} In order to study the sensitivity of the SIDIS background
subtraction to the Monte Carlo description, the effect of using unweighted
or weighted LEPTO samples for the background parameterisation is
investigated. Despite the different shapes of $E_{\textrm{miss}}$
distributions in the two cases, differences between background-corrected
asymmetries are very small as expected, because a bad parameterisation
of the background does not introduce a background
asymmetry. Additionally, for the 2007 set-up a second large Monte Carlo
background sample was generated using PYTHIA with default values of
parameters~\cite{Jegou:10}. The background shapes obtained
with weighted LEPTO and weighted PYTHIA are similar but in general the
latter results in about 10\,\% less background.  In most of the kinematic bins
the asymmetries are the same. The systematic uncertainty due to the
Monte Carlo description of the SIDIS background is negligible for the proton data and 
is estimated to be about
$0.04\,\sigma^{stat}$ for the deuteron data.

{\bf (f)} An important consistency test is the comparison of the mean
asymmetry values evaluated in bins of $x_{Bj}$, $Q^2$ and $p_T^2$, to
check if the assumption of Eq.~\eqref{eq:RA} holds after background
subtraction.
For the proton data, the mean
values of the asymmetries evaluated in bins of $x_{Bj}$ and $Q^2$ are compatible,
and the one evaluated in bins of $p_T^2$ agrees within about
$0.30\,\sigma^{stat}$; half of this
difference is taken into account as systematic uncertainty due to
background subtraction. 
For the deuteron data, the three mean asymmetry values are
in reasonable agreement. The small differences observed are similar as
in the case of the background non-corrected asymmetries and hence not
introduced by background subtraction. 

\begin{table}
\begin{center}
\caption{Estimates of systematic uncertainties of \AUT for proton and deuteron data. No value is quoted when the systematic uncertainty is negligibly small, below 0.01.}
\label{tab:sys_error}
\begin{tabular}{l c c}
\toprule
%source of systematic uncertainty & $~~~~~\sigma^{\mathit{sys}}/\sigma^{\mathit{stat}}\phantom{x}$\\
source of systematic uncertainty & 
$\sigma^{\mathit{sys}}/\sigma^{\mathit{stat}}$ &
$\sigma^{\mathit{sys}}/\sigma^{\mathit{stat}}$ \\
\midrule
& proton data & deuteron data\\
\midrule
(a) 1D binned estimator & 0.12 & 0.10 \\
(b) data stability    & --- & --- \\
(c) false asymmetries & 0.15 & 0.49 \\
(d) method of background subtraction & --- & --- \\
(e) MC dependence & --- & 0.04 \\
(f) compatibility after background subtraction in $x_{\mathit{Bj}}$, $Q^2$ and $p_T^2$ & 0.15 & ---\\
\midrule
total & 0.25 & 0.50  \\
\bottomrule
\end{tabular}
\end{center}
\end{table}

A summary of the systematic uncertainties (a)-(f) is given in
Table~\ref{tab:sys_error}. For each kinematic bin $i$, the total systematic
uncertainty of the measured asymmetry \AUT is
obtained as quadratic superposition of the sources (a) to (f). It equals 
$\sigma _i^{sys} = 0.25 \,
\sigma _i^{stat}$ for the proton and $\sigma _i^{sys} = 0.5 \, \sigma _i^{stat}$
for the deuteron data, where $\sigma _i^{stat}$
is the statistical uncertainty of the measured asymmetry \AUT in bin $i$.

Not listed in the table are the
scale uncertainties due to the relative
uncertainties of dilution factor and target polarisation. 
For the NH$_3$ target these are 2\% and 3\%, which leads by
quadratic superposition to a scale uncertainty of
0.036.
The analogous values for the $^6$LiD target are 2\% 
and 5\% ,
respectively, resulting in a scale uncertainty of
0.054. Note that these scale uncertainties are common 
for all measured asymmetries for a given target. 

\section{Results and comparison to model predictions}
 \label{Results}

\begin{figure}[htb]
 \begin{center}
 \epsfig{figure=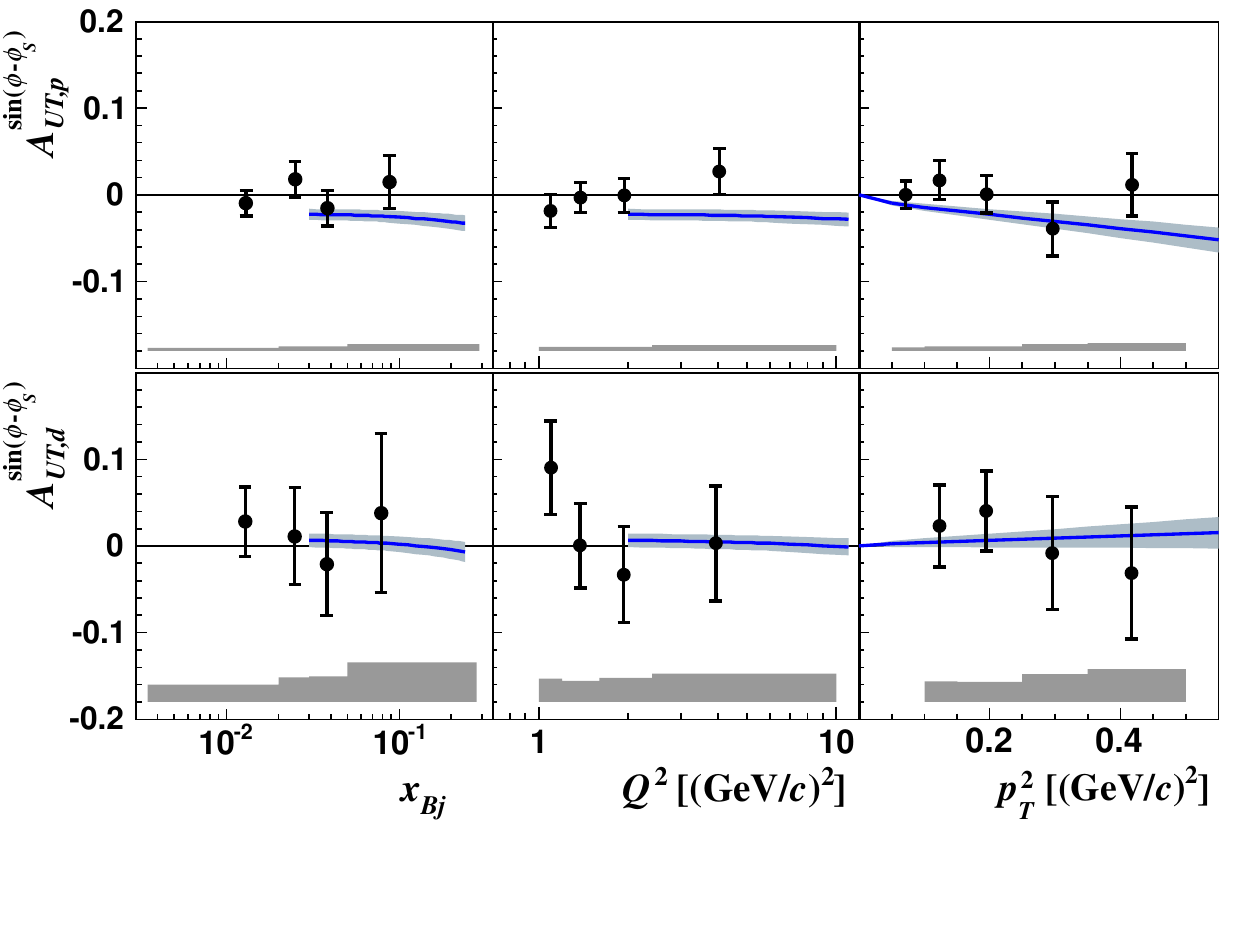,width=14cm,angle=0,clip,trim=0 35 0 0}
 \caption{Transverse target spin asymmetries $A_{UT}^{\rm{sin}(\phi-\phi_S)}$ measured on
proton (upper) and deuteron (lower) as a function of $x_{Bj}$, $Q^2$ 
and $p_T^2$. Error bars show statistical uncertainties, while the systematic
ones are represented by grey bands at the bottom. The curves show the predictions of the GPD 
model~\cite{gk3}
using the set of parameters called `variant 1'. They are calculated at 
$W$ = 8.1\,GeV/$c^2$ and $p_T^2$ = 0.2\,(GeV/$c$)$^2$ for the left and middle panels, and
at $W$ = 8.1\,GeV/$c^2$ and $Q^2$ = 2.2\,(GeV/$c$)$^2$ for the right panels. The theoretical error
bands reflect uncertainties of GPD parameterisations.
}
 \label{AUTproton-and-deuteron}
 \end{center}
\end{figure}

The transverse target spin asymmetries $A_{UT}^{\rm{sin}(\phi-\phi_S)}$
measured on proton and deuteron are shown in
Fig.~\ref{AUTproton-and-deuteron} as a function of $x_{Bj}$, $Q^2$ or
$p_T^2$, upon integrating over the two other variables. For both targets the asymmetries
are found to be small and consistent with zero within
statistical uncertainties.  Note that this is the first measurement of
\AUT for transversely polarised deuterons.
The numerical values for \AUT
are presented in Table~\ref{tab:asytable} for each $x_{Bj}$, $Q^2$ and
$p_T^2$ bin, together with statistical and systematic uncertainties.
Also, average values of kinematic variables for each bin are
given. Averaged over the COMPASS kinematic region, the values of \AUT
are $-0.002 \pm 0.010({\rm stat}) \pm 0.003({\rm sys})$ for the proton and $0.02 \pm 0.03({\rm stat}) \pm 0.02({\rm sys})$ for the deuteron.

% The systematic uncertainty is calculated as described in 
%Sec.~\ref{Systematics} and is represented by the bands at the bottom of each
%panel.
\begin{table}
\begin{center}
\caption{The transverse target spin asymmetries \AUT measured on
proton and deuteron  in bins of $Q^2$, $x_{Bj}$ and $p_T^2$. The systematic uncertainties are obtained using the values given in Table~\ref{tab:sys_error}. 
In addition, a scale uncertainty of 3.6\,\% (5.4\,\%) accounts 
for uncertainties in the determination of the target polarisation and
target dilution factor for proton (deuteron) data.}
\label{tab:asytable}
\begin{tabular}{c c c c c}
\toprule
\multicolumn{5}{c}{proton data}  \\
\cmidrule(r){2-5}
& $\langle Q^2 \rangle$ $(\rm{GeV}/\mathit{c})^2$& $\langle x_{\mathit{Bj}} \rangle$ & $\langle p_T^2 \rangle$ $(\rm{GeV}/\mathit{c})^2$& $ A_{UT,p}^{\sin(\phi-\phi_S)} \pm \sigma^{\mathit{stat}}\pm \sigma^{\mathit{sys}}$\\
\midrule
$Q^{2}$ bin $(\rm{GeV}/\mathit{c})^2$ & & & & \\
\midrule
$1.0 - 1.2\phantom{0}$   &  $1.1$  &  $0.019$ &  $0.18$  &  $-0.019 \pm 0.019 \pm 0.005$  \\
$1.2 - 1.6\phantom{0}$   &  $1.4$  &  $0.025$ &  $0.18$  &  $-0.003 \pm 0.018 \pm 0.004$  \\
$1.6 - 2.4\phantom{0}$   &  $1.9$  &  $0.035$ &  $0.18$  &  $-0.001 \pm 0.020 \pm 0.005$  \\
$2.4 - 10.0$  &  $4.0$  &  $0.076$ &  $0.19$  &  $\phantom{-}0.027 \pm 0.026 \pm 0.007$  \\
\midrule
$x_{\mathit{Bj}}$ bin & & & & \\
\midrule
$0.003 - 0.02$   &   $1.4$  &  $0.013$ &  $0.17$  &  $-0.010 \pm 0.015 \pm 0.004$  \\
$\phantom{0}0.02 - 0.03$   &   $1.6$  &  $0.025$ &  $0.18$  &  $\phantom{-}0.018 \pm 0.020 \pm 0.005$  \\
$\phantom{0}0.03 - 0.05$   &   $1.9$  &  $0.038$ &  $0.18$  &  $-0.015\pm 0.020 \pm 0.005$  \\
$\phantom{0}0.05 - 0.30$   &   $3.8$  &  $0.088$ &  $0.19$  &  $\phantom{-}0.015 \pm 0.031 \pm 0.008$  \\
\midrule
$p_{T}^{2}$ bin $(\rm{GeV}/\mathit{c})^2$ & & & & \\
\midrule
$0.05 - 0.10$    &  $2.1$  &  $0.037$ &  $0.07$  &  $\phantom{-}0.000 \pm 0.016 \pm 0.004$   \\
$0.10 - 0.15$    &  $2.1$  &  $0.039$ &  $0.12$  &  $\phantom{-}0.017 \pm 0.023 \pm 0.006$   \\
$0.15 - 0.25$    &  $2.2$  &  $0.040$ &  $0.20$  &  $\phantom{-}0.001 \pm 0.022 \pm 0.005$   \\
$0.25 - 0.35$    &  $2.2$  &  $0.042$ &  $0.30$  &  $-0.039 \pm 0.031 \pm 0.008$   \\
$0.35 - 0.50$    &  $2.3$  &  $0.043$ &  $0.42$  &  $\phantom{-}0.012 \pm 0.036 \pm 0.009$   \\
\midrule
\multicolumn{5}{c}{deuteron data}  \\
\cmidrule(r){2-5}
& $\langle Q^2 \rangle$ $(\rm{GeV}/\mathit{c})^2$& $\langle x_{\mathit{Bj}} \rangle$ & $\langle p_T^2 \rangle$ $(\rm{GeV}/\mathit{c})^2$& $ A_{UT,d}^{\sin(\phi-\phi_S)} \pm \sigma^{\mathit{stat}}\pm \sigma^{\mathit{sys}}$\\
\midrule
$Q^{2}$ bin $(\rm{GeV}/\mathit{c})^2$ & & & & \\
\midrule
$1.0 - 1.2\phantom{0}$   &  $1.1$  &  $0.018$ &  $0.23$  &  $\phantom{-}0.09~  \pm 0.05~ \pm 0.03~$ \\
$1.2 - 1.6\phantom{0}$   &  $1.4$  &  $0.023$ &  $0.23$  &  $\phantom{-}0.00~  \pm 0.05~ \pm 0.02~$ \\
$1.6 - 2.4\phantom{0}$   &  $1.9$  &  $0.031$ &  $0.23$  &  $-0.03~ \pm 0.06~ \pm 0.03~$ \\
$2.4 - 10.0$  &  $3.9$  &  $0.059$ &  $0.24$  &  $\phantom{-}0.00~  \pm 0.07~ \pm 0.03~$ \\
\midrule
$x_{\mathit{Bj}}$ bin & & & & \\
\midrule
$0.003 - 0.02$   &            $1.4$  &  $0.013$ &  $0.23$  &  $\phantom{-}0.03~  \pm 0.04~ \pm 0.02~$ \\
$\phantom{0}0.02 - 0.03$   &  $1.6$  &  $0.025$ &  $0.23$  &  $\phantom{-}0.01~  \pm 0.06~ \pm 0.03~$ \\
$\phantom{0}0.03 - 0.05$   &  $2.0$  &  $0.038$ &  $0.23$  &  $-0.02~ \pm 0.06~ \pm 0.03~$ \\
$\phantom{0}0.05 - 0.30$   &  $3.9$  &  $0.078$ &  $0.24$  &  $\phantom{-}0.04~  \pm 0.09~ \pm 0.05~$ \\
\midrule
$p_{T}^{2}$ bin $(\rm{GeV}/\mathit{c})^2$ & & & & \\
\midrule
$0.10 - 0.15$ &  $1.9$  &  $0.031$ &  $0.12$  &  $\phantom{-}0.02~  \pm 0.05~ \pm 0.02~$    \\
$0.15 - 0.25$ &  $2.0$  &  $0.031$ &  $0.19$  &  $\phantom{-}0.04~  \pm 0.05~ \pm 0.02~$    \\
$0.25 - 0.35$ &  $2.0$  &  $0.032$ &  $0.30$  &  $-0.01~ \pm 0.07~ \pm 0.03~$    \\
$0.35 - 0.50$ &  $2.1$  &  $0.033$ &  $0.42$  &  $-0.03~ \pm 0.08~ \pm 0.04`$    \\
\bottomrule
\end{tabular}
\end{center}
\end{table}
 
The results of a similar measurement of the asymmetry
$A_{UT}^{\rm{sin}(\phi-\phi_S)}$ for $\rho ^0$ production on the proton
target by the HERMES experiment~\cite{hermes1} are also consistent with
zero within total experimental uncertainties.  The separate asymmetries for longitudinally and
transversely polarised \r0 mesons were found by HERMES~\cite{hermes2} to
be consistent with zero as well.

Theoretical predictions for $A_{UT}^{\sin(\phi-\phi_S)}$ for $\rho ^0$
are given by the GPD model of Golo\-sko\-kov and Kroll~\cite{gk3}.
In this model, electroproduction of a light vector meson $V$ at small $x_{Bj}$
is analysed in the handbag approach, in which the amplitude of the process is a convolution of
GPDs with amplitudes for the partonic subprocesses $\gamma ^* q^f \rightarrow V q^f$ 
and $\gamma ^* g \rightarrow V g$. 
The partonic subprocess amplitudes, which comprise corresponding
hard scattering kernels and meson DAs, are calculated in the modified
perturbative approach where
the transverse momenta of quarks and antiquarks forming the vector meson
are retained and Sudakov suppressions are taken into account. The
model gives predictions for contributions from both longitudinal and
transverse virtual photons.

The predicted value of the proton asymmetry averaged over the COMPASS
kinematic region is about $-0.02$ and correspondingly about $-0.03$ for
HERMES. 
The comparison of the COMPASS results as a function of $x_{Bj}$, $Q^2$ and $p_T^2$
to the predictions of the model, shown in 
Fig.~\ref{AUTproton-and-deuteron}, indicates reasonable
agreement. The curves were obtained using the default version of the model (`variant 1'), with only 
contributions from valence
quark GPDs $E^u$ and $E^d$. The indicated theoretical error bands
reflect uncertainties in the GPD parameterisations. 

In order to investigate the role of gluons and sea quarks
the authors of the model consider two extreme cases for non-zero GPDs 
$E^g$ and $E^{sea}$.
They use either positive or negative $E^{sea}$ that saturates positivity bounds,
and $E^g$ that is constrained by a sum-rule for the second moments of GPDs $E$ of quarks and 
gluons~\cite{gk3}. Including GPDs $E^g$ and $E^{sea}$ has a very small effect on the predicted
values of \AUT, resulting in differences with respect to the default 
version which are significantly smaller than the theoretical uncertainties shown 
in Fig.~\ref{AUTproton-and-deuteron}.    
%The expected
%values of \AUT for incoherent exclusive \r0 production on the deuteron
%are even smaller~\cite{gk3} than those for the proton.

%$The small value of the asymmetry for \r0 is due to an
%approximate cancellation of two sizeable and comparable contributions of
%opposite sign from GPDs $E$ for valence $u$ and $d$ quarks, $E^u
%\approx - E^d$.  
The sensitivity of \AUT to the light quark GPDs,
$E^u$ and $E^d$, is different for the two targets. For the
proton these GPDs enter the amplitude as the sum $2/3\,E^u + 1/3\,E^d$.
For incoherent production on the nucleons of the deuteron,
assuming isospin invariance and neglecting nuclear effects, they
effectively contribute as $E^u + E^d$. In both cases, a small value of the 
asymmetry for \r0 is expected as $E^u$ and $E^d$ are similar in magnitude but
of opposite sign. 
%Thus in
%principle the measured asymmetry for the deuteron may provide
%complementary constraints for modelling the GPDs $E^f$.

%The Model Of Ref.~\Cite{Gk3} Predicts Sizeable Values Of \Aut
%A\-Sym\-Me\-Try For Other Vector Mesons, In Particular For Exclusive
%Production Of $\Omega $ Meson. In The Latter Case, Due To The Quark
%Flavour Content Of The Meson, The Gpd $E^D$ Contributes To The Amplitude
%For $\Omega$ Production With The Opposite Sign Than For \R0 Leading To
%An Enhancement Of The Asymmetry Value, Which Is Predicted To Be About
%-0.1. Measurements Of \Aut Asymmetry For This Channel, Although More
%Challenging Than That For $\Rho ^0$ Meson, Would Be Particularly
%Interesting.

In conclusion, the transverse target spin asymmetry \AUT for hard exclusive
 $\rho^0$ meson production was measured at COMPASS on the proton
and, for the first time, on the deuteron. The values of the asymmetry for
both targets are small and compatible with zero in a broad kinematic
range. They are compatible with the predictions of the GPD model of
Ref.~\cite{gk3}. The COMPASS proton results are in good agreement
with those obtained at HERMES, while they are more precise by a factor of about 3
 and cover a larger kinematic domain.

\noindent
We acknowledge the support of the CERN management and staff, as well
as the skills and efforts of the technicians of the collaborating
institutes. 
We also thank S.V. Goloskokov and 
P. Kroll for discussions of 
the results.

\end{document}